\documentclass[twocolumn,showpacs,preprintnumbers,amsmath,amssymbaps,prb,citeautoscript]{revtex4}

\usepackage{graphicx}
\usepackage{amssymb}
\usepackage{amsfonts}
\usepackage{subfig}
\usepackage{dcolumn}
\usepackage{bm}
\usepackage{float}
\usepackage{footnote}

\newcommand{\ket}[1]{\left| #1 \right>} 
\newcommand{\braket}[2]{\left< #1 \vphantom{#2} \right|
 \left. #2 \vphantom{#1} \right>} 

\newcommand{\req}[1]{Eq.~(\ref{#1})}

\newcommand{\rref}[1]{(\ref{#1})}
\newcommand{\reffig}[1]{Fig. \ref{#1}}

\numberwithin{equation}{section}

\begin{document}


\title{Three-Boson Bound States in Two Dimensions}

\author{Tianhao Ren}
 \email{tr2401@columbia.edu}
\author{Igor Aleiner}
 \email{aleiner@phys.columbia.edu}
\affiliation{Physics Department, Columbia University, New York, NY 10027, USA}

\date{\today}

\begin{abstract}
We investigate the possible existence of the bound state in the system of three bosons interacting with each other via zero-radius potentials in two dimensions (it can be atoms confined in two dimensions or tri-exciton states in heterostructures or dihalogenated materials). The bosons are classified in two species (a,b) such that a-a and b-b pairs repel each other and a-b attract each other, forming the two-particle bound state with binding energy $\epsilon_b^{(2)}$ (such as bi-exciton). We developed an efficient routine based on the proper choice of basis for analytic and numerical calculations. For zero-angular momentum we found the energies of the three-particle bound states $\epsilon^{(3)}_b$ for wide ranges of the scattering lengths, and found a universal curve of $\epsilon^{(3)}_b/\epsilon^{(2)}_b$ which depends only on the scattering lengths but not the microscopic details of the interactions.
\end{abstract}


\maketitle

\section{Introduction}
\label{sec:intro}
The quantum three-body problem was first solved by Skorniakov and Ter-Martirosian for three fermions in the zero-range-interaction limit \cite{STM}. The integral equation approach introduced by Skorniakov and Ter-Martirosian was then generalized to include finite and long range interactions by Faddeev \cite{Faddeevarticle}. It was recognized \cite{Danilov,Minlos} that the Skorniakov-Ter-Martirosian equation gives a spectrum that is not bounded from below. This pathology was then resolved by Efimov \cite{EFIMOV,EFIMOV1970563}, which gives a condensation of three-particle bound states at infinite scattering length. This trigged the fruitful field of Efimov physics in three dimensions \cite{,Braaten2006259,PhysRevD.5.1992,Hammer:2011kg,BEDAQUE1999444}. In two dimensions, the counterpart is studied for the case of three interacting bosons  \cite{PhysRevA.19.425,PhysRevA.22.28,0953-4075-44-20-205302,PhysRevA.85.025601,0953-4075-46-5-055301} and charged particles \cite{:/content/aip/journal/jcp/70/10/10.1063/1.437251,PhysRevLett.114.107401}. 

Our interest in few-body problems in two dimensions is trigged but not exhausted by the study of the many body physics in exciton Bose-Einstein condensates in GaAs-based quantum well structures\cite{RevModPhys.82.1489,nature,condensation,Deng199}. In these systems we have two kinds of bright excitons with spin projection $m=\pm 1$ to the structural axis, where the same spin projections repel each other and the opposite spin projections attract each other \cite{MILLER1985520,PhysRevB.25.6545,PhysRevB.38.3583}. In such systems with attractive inter-species coupling, formation of few-body bound states is the possible route to the instability of the condensates. This problem was addressed in three dimensions by Petrov \cite{PhysRevLett.115.155302}, whereas little is known about such instability in two dimensions\nocite{PhysRevLett.117.100401}.\footnote{At the time of writing this paper, Ref [25] has been extended to low dimensions, see Ref [26]} Investigation in our paper can be viewed as a first step towards quantitative understanding of the instabilities in two-component Bose-Einstein condensates in two dimensions, especially for the excitonic systems in quantum well structures.

In the literature, three-boson problems in two dimensions are only solved for the case with the same kind of interaction (either repulsive or attractive) between bosons \cite{PhysRevA.19.425,0953-4075-44-20-205302,PhysRevA.85.025601,0953-4075-46-5-055301,PhysRevA.73.032724}, To address the stability of the two dimensional system, we need to take into account both repulsive and attractive interactions, and for all possible scattering channels. We then consider three interacting bosons in two dimensions. The interactions between particles are short-ranged, and we model them as contact interaction with finite radius $r_0$. This is suitable for excitonic systems in quantum well structures, where the short-ranged exchange interaction is much stronger than the direct dipole-dipole interaction \cite{PhysRevB.59.10830}. We also make the choice that particle 1 and particle 2 are alike (species a) and repel each other; while particle 3 is different (species b) and attracts the other two. Then the Hamiltonian of the system under consideration is as follows (we choose the unit such that $m=\hbar=1$):
\small
\begin{equation}
\label{Ham}
\mathcal{H}=-\sum_{i=1,2,3}\frac{\nabla^2_i}{2}+\lambda_1\delta^2(\bm{r}_{12})-\lambda_2\left[\delta^2(\bm{r}_{13})+\delta^2(\bm{r}_{23})\right],
\end{equation}
\normalsize
where the two-dimensional $\delta$-function is understood to have a finite radius $r_0$. And $\lambda_1>0$ and $\lambda_2>0$ represent the repulsive and attractive couplings, whose low energy scattering lengths are denoted as $\alpha_<$ and $\alpha_>$ respectively:
\begin{equation}
\label{scatterlen}
\alpha_<=e^{\mathbb{C}}r_0\exp\left(-\frac{2\pi}{\lambda_1}\right)~; ~~~ \alpha_>=e^{\mathbb{C}}r_0\exp\left(\frac{2\pi}{\lambda_2}\right),
\end{equation}
where $\mathbb{C}=0.577\cdots$ is the Euler constant and we have the relation that $\alpha_<\ll r_0 \ll \alpha_>$.

Short-ranged interactions in two dimensions is well-known to present logarithmic poles in the low-energy scattering amplitude \cite{Landau,Adhikari,PhysRevA.64.012706}:
\begin{equation}
\label{eq:amp}
f_>(k)=-\frac{\sqrt{\pi/2k}}{\ln (2i/k\alpha_{>})}; ~~~ f_<(k)=-\frac{\sqrt{\pi/2k}}{\ln (2i/k\alpha_{<})},
\end{equation}
where $k=\sqrt{2\mu\epsilon}$ is the momentum associated with the two-particle energy, and $\mu$ is the reduced mass which in our case equals to $1/2$. The expression (\ref{scatterlen}) gives the two-particle binding energy $\epsilon^{(2)}_b$ for the attractive potential as follows (at such energy $f_>(i\sqrt{\epsilon^{(2)}_b})\rightarrow \infty$):
\begin{equation}
\epsilon_b^{(2)}=\frac{4}{\alpha^2_{>}}.
\end{equation}
The corresponding pole for the repulsive potential occurs at momentum $|k|\gg 1/r_0$, which is beyond the logarithmic pole approximation, and must be disregarded in the calculation as a spurious solution. 

The purpose of this paper is to analyze the three-particle bound state energies $\epsilon^{(3)}_b$ as functions of the scattering lengths $\alpha_>$ and $\alpha_<$. The remainder of the paper is organized as follows. In Sec. \ref{sec:formalism} we introduce the parameterization scheme of the problem, and give the formal solution to the resulting one-dimensional Schrodinger equation via a boundary-matching-matrix technique. We also introduce a convenient running basis to the problem, which is suited for numerical implementations. In Sec. \ref{sec:eigenstates} we give out the explicit solutions for zero and nonzero angular momentum separately. Large scale behaviors are analyzed analytically and three-particle binding energies are calculated numerically. Finally in Sec. \ref{sec:conclusion} we summarize the results and compare our methods with existing ones. Technical details are relegated to the Appendices.

\section{Formalism}
\label{sec:formalism}

\subsection{Parameterization of the Configuration Space}
\label{sec:formalismA}
For the configuration space of the system under consideration, we use the Faddeev parameterization \cite{Faddeev} 
\begin{equation}
\bm{r}_{12}=\bm{r}_1-\bm{r}_2, ~~~ \bm{\rho}_3=(\bm{r}_1+\bm{r}_2-2\bm{r}_3)/\sqrt{3}.
\end{equation}
After that, we perform the usual separation of radial and angular parts of the four dimensional vector $(\bm{r}_{12},\bm{\rho}_3)^T$:
\begin{equation}
\begin{pmatrix}\bm{r}_{12} \\ \bm{\rho}_3\end{pmatrix}=r\bm{N}, ~~~ \bm{N}^2=1.
\end{equation}
This spherical separation enables us to assign a discrete set of angular level labels $j$ for the wave function $\bm{\Phi}=(\Phi_0,\Phi_1,\cdots)^T$, due to the fact that the angular momentum operator is compact \cite{Edmonds,Dong}.

Usually, the angular part of four dimensional vector is represented in terms of hyperspherical coordinates in the literature \cite{Letz1975,Nielsen1999,Nielsen2001373,PhysRevA.74.042506,PhysRevLett.112.103201,PhysRevA.90.042707,PhysRevA.91.062710,PhysRevA.82.022706,PhysRevA.93.012511}, but the resulting algorithms have slow convergence and the number of states scales as the square of the number of levels included. Here we adopt the Hopf coordinates, which gives faster convergence and number of states proportional to the number of levels included (see Appendix \ref{app:Hopf} ):
\begin{equation}
\bm{N}=\begin{pmatrix}
\sqrt{\frac{1-x}{2}}\cos\phi_1 \\ \sqrt{\frac{1-x}{2}}\sin\phi_1 \\\sqrt{\frac{1+x}{2}}\cos\phi_2 \\ \sqrt{\frac{1+x}{2}}\sin\phi_2
\end{pmatrix}. \label{Hopf}
\end{equation}
Substituting the above parameterization of the configuration space into \req{Ham}, we will get the following one-dimensional matrix Schrodinger equation:
\begin{equation}
\label{oned}
\mathcal{H}\bm{\Phi}=\Big{[}-\frac{1}{r^3}\frac{\partial}{\partial r}r^3\frac{\partial}{\partial r}+\frac{\hat{U}(r)}{r^2}\Big{]}\bm{\Phi}=\epsilon\bm{\Phi},
\end{equation}
where the effective potential operator $\hat{U}(r)$ is a sum of angular momentum operator and the interaction term:
\begin{equation}
\label{eq:effpotential}
\hat{U}(r)=4\hat{L}^2+r^2\hat{V}_r(\bm{n}).
\end{equation}
The angular momentum operator under Hopf coordinates has the following form:
\begin{equation}
\label{Laplacian}
\hat{L}^2=-\frac{\partial}{\partial x}(1-x^2)\frac{\partial}{\partial x}-\frac{\partial^2_{\phi_1}}{2(1-x)}-\frac{\partial^2_{\phi_2}}{2(1+x)},
\end{equation}
and the interaction term can be written in the following form showing explicitly the scale dependence (see Appendix \ref{app:Hopf}):
\begin{equation}
\hat{V}_r(\bm{n})=\frac{2}{\pi r^2}\sum_{i=1,2,3}\mu_i\delta_r(1-\bm{n}\cdot\bm{n}_i),
\end{equation}
where $\mu_1=\lambda_1$ and $\mu_{2,3}=-\lambda_2$ are the repulsive and attractive coupling constants respectively; the scale dependent $\delta$-function is defined as $\delta_r(x)=\delta(x-2r^2_0/r^2)$, which takes care of the finite radius. The configuration space is projected onto the three-dimensional unit sphere ($\phi=\phi_1-\phi_2$):
\begin{subequations}
\begin{align}
& \bm{n}=(\sqrt{1-x^2}\cos\phi,\sqrt{1-x^2}\sin\phi,x), \label{unit}\\
& \bm{n}_1=(0,0,1), ~~~ \bm{n}_{2,3}=(\pm\frac{\sqrt{3}}{2},0,-\frac{1}{2}).
\end{align}
\end{subequations}
The total angular momentum $m$ is a good quantum number because its corresponding operator commutes with the Hamiltonian:
\begin{equation}
\left[ -i\left(\frac{\partial}{\partial \phi_1}+\frac{\partial}{\partial \phi_2}\right),\mathcal{H}  \right]=0.
\end{equation}
For each $m$ the Hilbert state is characterized by the three-dimensional angular momentum $j$ (integer for even $m$ and half-integer for odd $m$). The eigenvalue of $\hat{L}^2$ is of order $j^2$ and the degeneracy of each level is $(2j+1)$. Also the bosonic symmetry of the system require the following symmetry property of the eigenfunction $\bm{\Phi}(\bm{n})$:
\begin{equation}
\bm{\Phi}(n_x,n_y,n_z)=\bm{\Phi}(-n_x,n_y,n_z).
\end{equation}

The interaction term makes the states deviate from free motion. There are three $\delta$-functions in total, thus at most three states are affected for each level $j$. Because we are considering a bosonic system, only symmetric states are physical, which leaves us at most two affected states for each level $j$,  all the other states can be ignored because they belong to the space orthogonal to the possible physical bound states. Hopf coordinates is such a choice that enables us to identify the relevant states directly, instead of representing them as a sum of many hyperspherical harmonics (for more detail, see Appendix \ref{app:Hopf}).

\subsection{Solution of the One-Dimensional Schrodinger Equation}
\label{sec:Levinson}
After the effective potential operator $\hat{U}(r)$ is obtained, we are left with the problem of solving the one-dimensional matrix Schrodinger equation \rref{oned}. Naive approach to this radial equation is to numerically solve \req{oned} by limiting the basis to $N$ functions, but it is practically inaccessible due to the exponential instability of the wave function even if one of the $N$ boundary conditions or energies is not chosen correctly. Thus we choose another approach \cite{PhysRevLett.114.107401}, converting the Schrodinger equation \rref{oned} into a first order nonlinear differential equation for the boundary-matching-matrix $\hat{\Lambda}(r)$ defined as follows:
\begin{equation}
\label{boundary}
r\frac{d\bm{\Phi}}{dr} \Big{|}_{r=R}=-\hat{\Lambda}(R)\bm{\Phi}(R).
\end{equation}
Then the differential equation of $\hat{\Lambda}(r)$ is obtained by requiring the invariance of  \req{boundary} with respect to length scale $R$:
\begin{equation}
\label{diff}
\left[\frac{d\bm{\Phi}}{dr}+r\frac{d^2\bm{\Phi}}{dr^2}\right]_{r=R}=-\frac{d\hat{\Lambda}}{dR}\bm{\Phi}(R)-\hat{\Lambda}(R)\frac{d\bm{\Phi}}{dr}\Big{|}_{r=R}.
\end{equation}
From the Schrodinger equation \rref{oned} we have
\begin{equation}
\frac{d^2\bm{\Phi}}{dr^2}=-\frac{3}{r}\frac{d\bm{\Phi}}{dr}+\Big{(}\frac{\hat{U}}{r^2}-\epsilon\Big{)}\bm{\Phi}.
\end{equation}
Substitute this back into \req{diff} and multiply both sides by $r=R$, then we obtain
\begin{equation}
\left[\left(\hat{\Lambda}(r)-2\right)r\frac{d\bm{\Phi}}{dr}+(\hat{U}-r^2\epsilon)\bm{\Phi}\right]_{r=R}=-R\frac{d\hat{\Lambda}}{dR}\bm{\Phi}(R).
\end{equation}
Finally refer back to definition of $\hat{\Lambda}$, which is \req{boundary}, and we obtain the radial renormalization equation:
\begin{equation}
\label{renorm1}
\frac{d\hat{\Lambda}}{d\ln r}=r^2\epsilon-\hat{U}(r)-2\hat{\Lambda}+\hat{\Lambda}^2.
\end{equation}
The advantage of the boundary-matching-matrix method is its numerical stability, meaning that even if the original wave function is subject to exponential growth with respect to $r$, our newly defined  matrix $\Lambda(r)$ is subject to at most linear growth:
\begin{equation}
||\bm{\Phi}(r)||\sim \exp(r) \Rightarrow ||\hat{\Lambda}(r)|| \lesssim r.
\end{equation}
The initial condition for \req{renorm1} is obtained as a solution in the region $r_0\ll r\ll 1$, where only kinetic energy is important:
\begin{equation}
\label{initialcon}
\frac{d\hat{\Lambda}}{d\ln r}\Big{|}_{r\rightarrow 0}=0 ~~~ \hat{\Lambda}(r\rightarrow 0)=\Big{(}1-\sqrt{4\hat{L}^2+1}\Big{)},
\end{equation}
then the initial matrix $\hat{\Lambda}(r\rightarrow0)$ is diagonal:
\begin{equation}
\Lambda_{ij}(r\rightarrow 0)=-2l_i\delta_{ij},
\end{equation}
where $l_i(l_i+1)$ is the eigenvalue of angular momentum operator $\hat{L}^2$ for level $i$.

\begin{figure}
\includegraphics[scale=0.15]{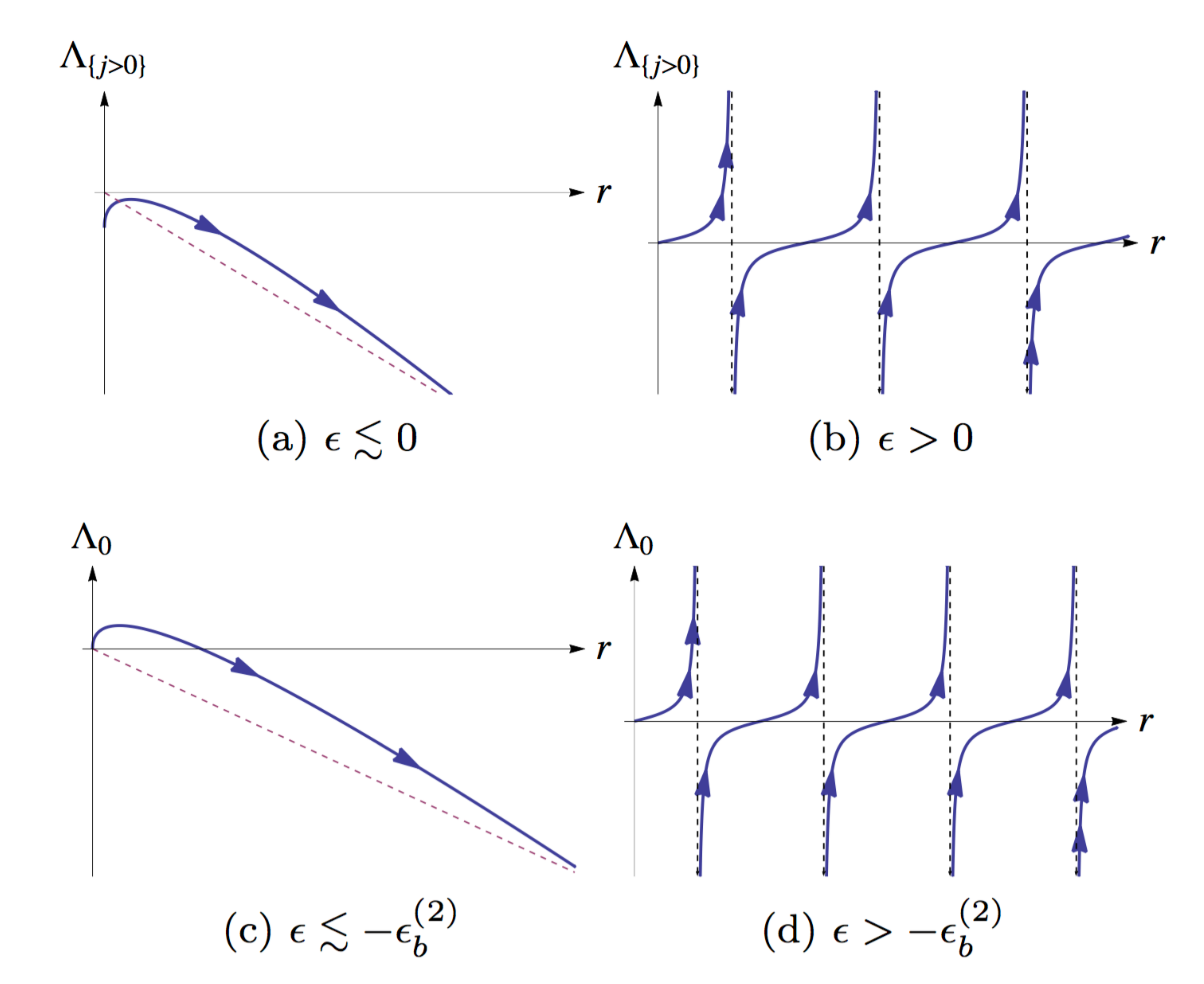}
\captionsetup{justification=raggedright}
\caption{\label{fig:large} Schematic diagram for large scale behavior of \req{renorm1}, where (a) and (b) are shown for levels $j\neq 0$, (c) and (d) are shown for the lowest level $j=0$. Left is shown for energy slightly below (a) zero for $j\neq 0$ (c) $-\epsilon^{(2)}_b$ for $j=0$. Right is shown for energy well above (b) zero for $j\neq 0$ (d) $-\epsilon^{(2)}_b$ for $j=0$.} 
\end{figure}

The large scale ($r\rightarrow \infty$) behavior of \req{renorm1} is determined by setting $U_{ij}(r)\simeq -r^2\epsilon^{(2)}_b\delta_{i0}\delta_{j0}$, where $\epsilon^{(2)}_b$ is the two-particle threshold in application to the Hamiltonian defined in \req{oned}. The equation has a stable trajectory for $\epsilon<0$ and $j\neq 0$:
\begin{equation}
\label{stabletrajectory1}
\Lambda_{ij}=-\delta_{ij}\sqrt{|\epsilon|}r ~~~ (j\neq 0).
\end{equation}
While for $\epsilon>0$ and $j\neq 0$, the trajectory shows periodic divergence jumps, typical for a spherical wave. For the lowest level $j=0$, there are also two situations: If $\epsilon<-\epsilon^{(2)}_b$, then the solution will also goes to a stable trajectory as
\begin{equation}
\label{stabletrajectory2}
\Lambda_0=-\sqrt{|\epsilon+\epsilon^{(2)}_b|}r.
\end{equation} 
If $\epsilon>-\epsilon^{(2)}_b$, the solution again corresponds to a spherical wave, which has periodical divergence jumps at the position that are zeros of the wave function (see \reffig{fig:large}). These divergent solutions actually form the continuum of the states of one bound biexciton and one exciton far away.

\begin{figure}
\includegraphics[scale=0.15]{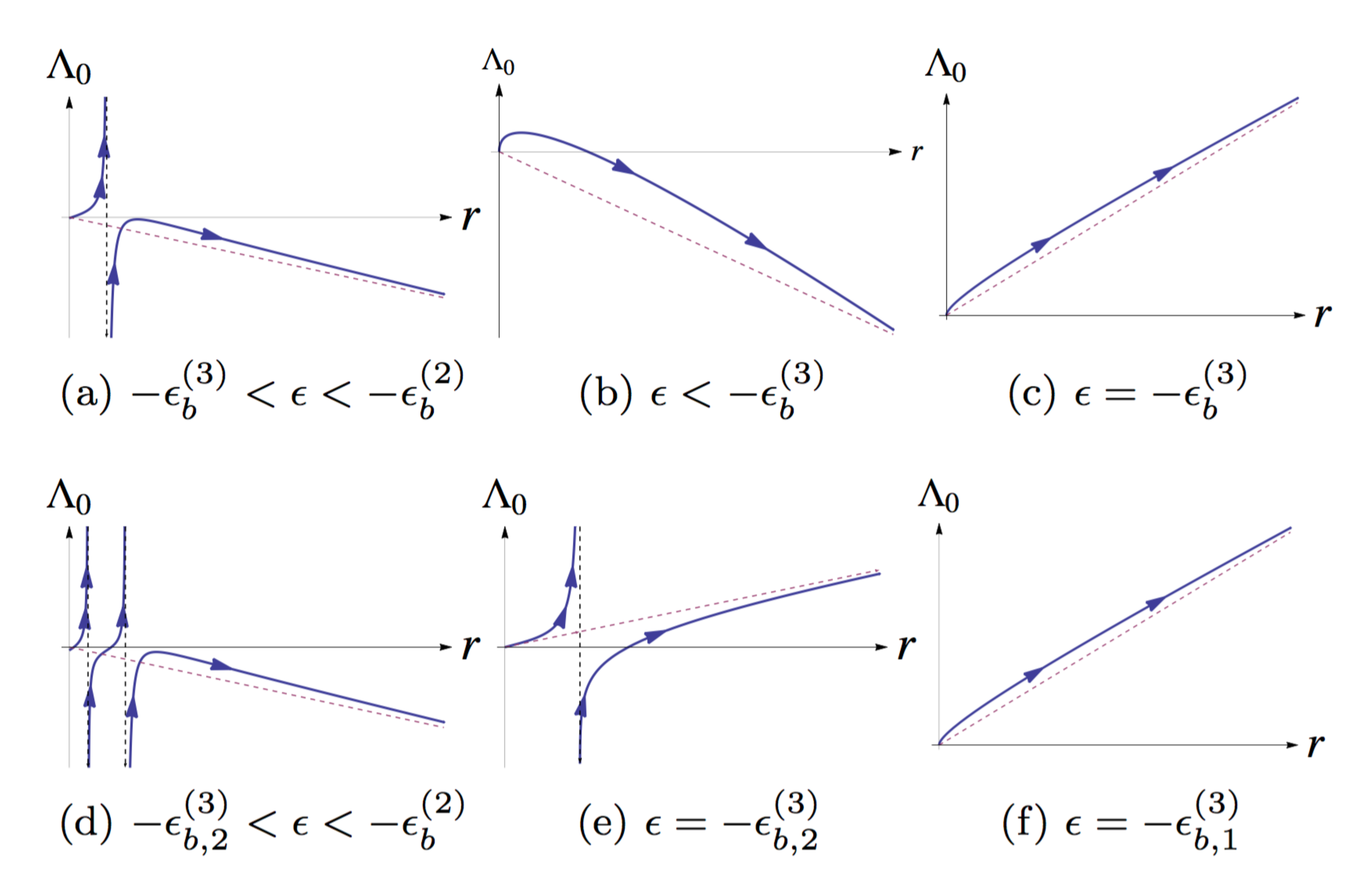}
\captionsetup{justification=raggedright}
\caption{\label{fig:intermediate} Schematic diagram for intermediate scale behavior of \req{renorm1}. Above: There is only one bound state. Below: There is two bound states, where we have $-\epsilon^{(3)}_{b,1}<-\epsilon^{(3)}_{b,2}$. Note that if only (c) or (f) is realized, bound state does not exist.}
\end{figure}

In the intermediate region, we solve for the possible three-particle bound states. The bound state is determined by the way $\Lambda_0$ approaches the stable trajectory defined in \req{stabletrajectory2}, and two typical situations are shown in \reffig{fig:intermediate}: (1) There is only one three-particle bound state with binding energy $\epsilon^{(3)}_b$. If the energy is between the three-particle binding energy $-\epsilon^{(3)}_b$ and the two-particle threshold $-\epsilon^{(2)}_b$,  the evolution of $\Lambda_0$ will show a single jump before attracted to the stable trajectory; If the energy is smaller than $-\epsilon^{(3)}_b$, $\Lambda_0$ will be directly attracted to the stable trajectory; The evolution of $\Lambda_0$ will diverge only when the energy is tuned exactly at the three-particle binding energy. (2) There are two three-particle bound states with binding energies $-\epsilon^{(3)}_{b,1}<-\epsilon^{(3)}_{b,2}$. The evolution of $\Lambda_0$ with different energies is similar to the previous case, but it will show two jumps before attracted to the stable trajectory if the energy is tuned to lie between $-\epsilon^{(3)}_{b,2}$ and $-\epsilon^{(2)}_b$. Following this line of reasoning, we can see the fact that the number of three-particle bound states is determined by the number of infinite jumps of $\Lambda_0$ at $\epsilon\lesssim -\epsilon^{(2)}_b$, which is exactly the content of the Levinson theorem \cite{Lin:1998pa,atomic}.\\

\subsection{Running Basis}
\label{sec:Run}
Sometimes, the following running basis that diagonalizes matrix $\hat{U}(r)$ is most convenient for both analytic and numerical calculations:
\begin{equation}
\hat{O}=\Big{(}\ket{\chi_0},\ket{\chi_1},\cdots \Big{)}, ~~~ \hat{U}(r)\ket{\chi_j}=u_j(r)\ket{\chi_j},
\end{equation}
where $\ket{\chi_j}$ is the angular part of the $j$-th component of the normalized wave function vector $\bm{\Phi}(r)$, whose expression will be derived latter in Sec. \ref{sec:eigenstates} via the Green's function method. This set of basis is called running basis because it changes with the length scale $r$. Then we do an unitary transformation to bring \req{renorm1} to the running basis:
\begin{subequations}
\begin{align}
& \hat{U}=\hat{O}\tilde{U}\hat{O}^{-1}, ~~~ \tilde{U}_{ij}=\delta_{ij}u_i(r),\\
& \hat{\Lambda}=\hat{O}\tilde{\Lambda}\hat{O}^{-1},
\end{align}
\end{subequations}
then the radial renormalization equation under the running basis reads (hereinafter we will drop the tilde symbol for simplicity):
\begin{equation}
\label{renorm2}
\frac{d\hat{\Lambda}}{d\ln r}+[\hat{\Lambda},\hat{D}]=r^2\epsilon-\hat{U}-2\hat{\Lambda}+\hat{\Lambda}^2,
\end{equation}
where the anti-symmetric matrix $\hat{D}$ is the Berry connection:
\begin{equation}
\label{Dij}
\hat{D}=\frac{d\hat{O}^{-1}}{d\ln r}\hat{O}, ~~\mathrm{i.e.}~~ D_{ij}=-D_{ji}=\braket{\frac{d\chi_i}{d \ln r}}{\chi_j}.
\end{equation}

It is very tempting (at least at large length scales) to neglect $\hat{D}$ altogether, which corresponds to the adiabatic approximation with a diagonal matrix $\hat{\Lambda}$. However it is not correct because of the following reason. Consider the lowest order correction $\delta\Lambda_{ij}$ to the adiabatic result $\Lambda^{(0)}_{ij}=\Lambda_{i}\delta_{ij}$ for the lowest level $(i=0)$, then the renormalization group equation for $\Lambda_0(r)$ reads:
\begin{equation}
\label{lowrenorm}
\begin{split}
\frac{d \Lambda_0}{d\ln r}=& r^2\epsilon-u_0(r)-2\Lambda_0+\Lambda^2_0\\
&-\sum_{j\neq 0}\left(\delta\Lambda_{0j}D_{j0}+D_{0j}\delta\Lambda_{j0}\right)+\sum_{j\neq 0}\delta\Lambda_{0j}\delta\Lambda_{j0},
\end{split}
\end{equation}
where $\delta\Lambda_{0j}$ can be obtained from the first order correction to the adiabatic approximation of \req{renorm2}:
\begin{equation}
\Lambda_{0}D_{0j}-D_{0j}\Lambda_{j}=-2\delta\Lambda_{0j}+\Lambda_0\delta\Lambda_{0j}+\delta\Lambda_{0j}\Lambda_j,
\end{equation}
which gives us the expression for $\delta\Lambda_{0j}$ as:
\begin{equation}
\delta \Lambda_{0j}=\frac{\Lambda_0-\Lambda_j}{\Lambda_0+\Lambda_j-2}D_{0j}. \label{d0j}
\end{equation}
Substituting the expression for $\delta\Lambda_{0j}$ into \req{lowrenorm} and using the anti-symmetry of the Berry connection $\hat{D}$, we finally obtain:
\begin{equation}
\label{deltau}
\begin{split}
\frac{d \Lambda_0}{d\ln r}=&\left[r^2\epsilon-\Big{(}u_0(r)+\sum_{j\neq 0}|D_{0j}|^2\Big{)} \right]-2\Lambda_0+\Lambda^2_0\\
& +\sum_{j\neq 0}|D_{0j}|^2\left[ \frac{2\Lambda_0-2}{\Lambda_0+\Lambda_j-2} \right]^2.
\end{split}
\end{equation}
The large scale behavior of the solution is determined by the following quantity:
\begin{equation}
\label{gamma}
\lim_{r\rightarrow \infty}\left[r^2\epsilon-\Big{(}u_0(r)+\sum_{j\neq 0}|D_{0j}|^2\Big{)}\right]_{\epsilon=-\epsilon^{(2)}_b}\equiv \gamma.
\end{equation}
If $\gamma>1$, the solution is unstable at $\epsilon=-\epsilon^{(2)}_b$, it has infinite number of jumps, which would correspond to infinite number of three-particle bound states. If $\gamma<1$, the solution is stable, it corresponds to the power law decay of the wave function. Only for the marginal value $\gamma=1$,  should the situation correspond to the non-interacting particle (one exciton and one biexciton) in two dimensions. On the physical ground we should have $\gamma=1$, thus it is important to check for the consistency by direct calculation of the quantity $\gamma$, taking into account the Berry connection as in \req{gamma}. We will show this calculation in later sections, see \req{eq326}.

In summary, we have shown in this section that the running basis is a convenient choice, whose leading order is the usual adiabatic approximation \cite{PhysRevA.82.022706,PhysRevA.90.042707} and the correction to it is the Berry connection. we have also argued that the Berry connection must be included for physically consistent calculation, thus we will use the exact formalism in our numerical calculation shown later.

\section{Eigenstates and Eigenvalue of Operator $\hat{U}(r)$}
\label{sec:eigenstates}

To obtain the full solution of the problem, we need to solve for the eigenvalues and eigenfunctions (which define our running basis) of operator $\hat{U}(r)$. We define the following Green's function for the angular Laplacian near pole $\bm{n}'$:
\begin{equation}
\label{Green}
\Big{[}4\hat{L}^2-u_j(r)\Big{]}G_j(\bm{n},\bm{n}')=\frac{2}{\pi}\delta_r(1-\bm{n}\cdot\bm{n}').
\end{equation}

We first solve the Green's function with $\bm{n}'$ along the north pole ($\bm{n}'=\bm{n}_1$), then perform $SO(4)$ rotations to obtain the Green's functions near the other two poles. After that we can use the obtained Green's function to make the following ansatz for eigenfunctions of operator $\hat{U}(r)$, taking into account the bosonic symmetry:
\small
\begin{subequations}
\begin{align}
& \chi_j(\bm{n})=\alpha_jG_{j}(\bm{n},\bm{n}_1)+\beta_j[G_{j}(\bm{n},\bm{n}_2)+G_{j}(\bm{n},\bm{n}_3)], \label{eigenstate1} \\
& \hat{U}(r)\chi_j(\bm{n})=u_j(r)\chi_j(\bm{n}). \label{eigenstate2}
\end{align}
\end{subequations}
\normalsize
Once the eigen-problem of operator $\hat{U}(r)$ is solved, then it is straightforward to solve \req{renorm2} analytically or numerically.\\
The solution of \req{Green} for $\bm{n}'=\bm{n}_1$ can be variable-separated:
\small
\begin{subequations}
\begin{align}
& G_j(\bm{n},\bm{n}_1)=G_j(x)e^{im_1\phi_1+im_2\phi_2},\\
& \Big{[}4\hat{Q}_{m_1,m_2}-u_j(r)\Big{]}G_j(x)=\frac{2}{\pi}\delta_r(1-x),\\
& \hat{Q}_{m_1,m_2}=-\frac{\partial}{\partial x}(1-x^2)\frac{\partial}{\partial x}+\frac{m^2_1}{2(1-x)}+\frac{m^2_2}{2(1+x)}.
\end{align}
\end{subequations}
\normalsize
As discussed in Sec.\ref{sec:formalismA}, total angular momentum $m=m_1+m_2$ is a good quantum number, therefore we can consider different angular momentum separately. We will first discuss the case with zero angular momentum, where three-particle bound state is possible; then we will show that no three-particle bound state exists for non-zero angular momentum.

\subsection{Zero Angular Momentum: Analytics}
In this section, we will analyze the large scale behavior of the case with zero angular momentum. It can be solved in two limiting cases, one of which agrees with the perturbative result and the other one shows the importance of including the Berry connection for the system to have physical marginal value $\gamma=1$.

For zero angular momentum we are dealing with the following Green's function:
\begin{equation}
\left[-4\frac{\partial}{\partial x}(1-x^2)\frac{\partial}{\partial x}-u_j(r)\right]G_j(x)=\frac{2}{\pi}\delta_r(1-x).
\end{equation}
This is just the Legendre equation of degree $\nu_j$ (except near point $x=1$) if we make the following substitution:
\begin{equation}
\label{nu}
u_j=4\nu_j(\nu_j+1).
\end{equation} 
Then the solution can be obtained by comparing the singularities \cite{Tables}  near point $x=1$, which gives us the following expression for the Green's function (here we use subscript $\nu_j$ instead of $j$ for Green's function to emphasize the dependence on degree $\nu_j$):
\begin{equation}
\label{Greenfunc}
G_{\nu_j}(x)=\frac{1}{4\cos\left[(\nu_j+1/2)\pi\right]}P_{\nu_j}(-x),
\end{equation}
and it is regularized at point $x=1$ by the finite radius $r_0$:
\begin{equation}
\label{reg}
G_{\nu_j}(1)=\frac{1}{4\pi }\left[\ln\frac{16}{\delta}-\Psi(-\nu_j)-\Psi(\nu_j+1)+2\Psi\left(\frac{1}{2}\right)\right],
\end{equation}
where $\delta=r^2_0/r^2$ and $\Psi(x)$ is the digamma function.\\
In the sector of zero angular momentum, only scalar-like combinations will enter the wave function, thus the specification of \req{eigenstate1} to zero angular momentum is
\begin{equation}
\label{ansatz}
\chi_j(\bm{n})=\alpha_jG_{\nu_j}(\bm{n}\cdot\bm{n}_1)+\beta_j\left[G_{\nu_j}(\bm{n}\cdot\bm{n}_2)+G_{\nu_j}(\bm{n}\cdot\bm{n}_3)\right].
\end{equation}
Substitute this ansatz into \req{eigenstate2}, we will obtain the following constraints on the coefficients:
\begin{equation}
\label{coeff}
\begin{pmatrix} \frac{1}{\lambda_1}+G_{\nu_j}(1); & 2G_{\nu_j}\left(-\frac{1}{2}\right) \\ 
-G_{\nu_j}\left(-\frac{1}{2}\right); & \frac{1}{\lambda_2}-G_{\nu_j}(1)-G_{\nu_j}\left(-\frac{1}{2}\right) \end{pmatrix}\begin{pmatrix}\alpha_j \\ \beta_j\end{pmatrix}=0.
\end{equation}
By setting the determinant to zero we obtain the equation of the spectrum:
\begin{widetext}
\begin{equation}
\label{spectrum1}
\left[\ln\frac{r}{\alpha_<}-F(\nu_j)+2\pi G_{\nu_j}\left(-\frac{1}{2}\right) \right]\left[\ln\frac{r}{\alpha_>}-F(\nu_j)+4\pi G_{\nu_j}\left(-\frac{1}{2}\right) \right]=2\left[2\pi G_{\nu_j}\left(-\frac{1}{2}\right)\right]^2,
\end{equation}
\end{widetext}
where the function $F(\nu_j)$ is defined as:
\begin{equation}
F(\nu_j)=\frac{1}{2}\Big{[}\Psi(-\nu_j)+\Psi(\nu_j+1) \Big{]}+2\pi G_{\nu_j}\left(-\frac{1}{2}\right).
\end{equation}
Here $\alpha_{>,<}$ are the scattering lengths for attractive and repulsive coupling respectively, see \req{scatterlen}.

The solution to the equation of spectrum can be solved analytically in the following two limiting cases: $u_0\rightarrow 0$ and $|u_0|=-u_0\rightarrow \infty$; while for general cases we will solve it numerically. In case of $u_0\rightarrow 0$, we have $u_0\sim 4\nu_0\rightarrow 0$ from \req{nu}. We first rewrite \req{spectrum1} into a more convenient form:
\begin{equation}
\frac{2}{\ln \frac{r}{\alpha_>}-F(\nu_j)}+\frac{1}{\ln \frac{r}{\alpha_<}-F(\nu_j)}=-\frac{1}{2\pi G_{\nu_j}(-\frac{1}{2})},\label{spectrum2}
\end{equation}
then we substitute the following behaviors for relevant functions into the above equation:
\begin{subequations}
\begin{align}
& F(\nu_0\rightarrow 0)\sim -\mathbb{C}-\ln\frac{\sqrt{3}}{2}+O(\nu_0),\\
& 2\pi G_{\nu_0}\left(-\frac{1}{2}\right)\Big{|}_{\nu_0\rightarrow 0}\sim -\frac{1}{2\nu_0}-\ln \frac{\sqrt{3}}{2}+O(\nu_0).
\end{align}
\end{subequations}
Finally, we obtain the following solution:
\begin{equation}
\frac{u_0}{2}\sim 2\nu_0=\frac{2}{\ln \frac{r}{\alpha_>}-F(0)}+\frac{1}{\ln \frac{r}{\alpha_<}-F(0)},
\end{equation}
that is just the perturbative result of the effective potential $u_0(r)$.

In the case of $|u_0|\rightarrow \infty$, we have the following asymptotic behaviors:
\begin{subequations}
\begin{align}
& \nu_0=-\frac{1}{2}+i\lambda, ~~~ \lambda=\frac{1}{2}\sqrt{|u_0+1|}\rightarrow \infty, \\
& F(\nu_0)\sim \ln \lambda-\frac{1}{24\lambda^2}+O\left(\frac{1}{\lambda^3}\right),\\
& G_{\nu_0}\left(-\frac{1}{2}\right)\sim \frac{1}{2\sqrt{\pi}3^{1/4}}\exp\left(-\frac{2\pi}{3}\lambda\right),
\end{align}
\end{subequations}
then using \req{spectrum1} we obtain the following solution:
\begin{equation}
\label{eqn317}
\ln\frac{r}{\alpha_>}= \frac{1}{2}\ln |u_0|-\ln 2-\frac{2}{3|u_0|}.
\end{equation}
The other solution associated with $\alpha_<$ corresponds to the spurious state discussed previously in the introduction section and should be dropped. Solving \req{eqn317} iteratively we will obtain the large scale behavior of the effective potential:
\begin{equation}
\label{adiabatic}
u_0(r\rightarrow \infty)=-r^2\epsilon^{(2)}_b-4/3+O(r^{-2}),
\end{equation}
where $\epsilon^{(2)}_b=4/\alpha^2_>$ is the two-particle threshold. According to the discussion at the end of Sec. \ref{sec:Run}, this result will give us $\gamma=\frac{4}{3}>1$ in the adiabatic approximation, which shows the necessity of including the Berry connection $D_{ij}$ [see \req{Dij}].

\begin{figure}
\includegraphics[width=7.5cm]{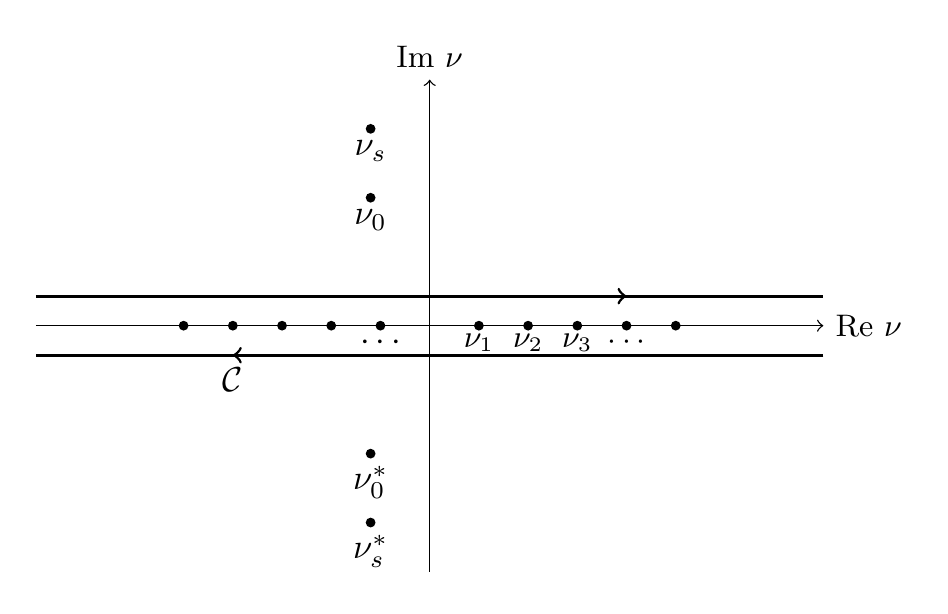}
\captionsetup{justification=raggedright}
\caption{\label{contour}Integration contour for the calculation of $\Delta u_0$. The contour is along real axis, where the first order poles reside. There are four extra poles off the real axis, which correspond to true bound state ($\nu_0$) and spurious bound state ($\nu_s$) respectively. The physical meaning of true bound state and spurious bound state is discussed at the end of the Sec. \ref{sec:intro}}
\end{figure}

The integral expression for the Berry connection $D_{ij}$ is:
\begin{equation}
D_{ij}=\frac{1}{8\pi^2\sqrt{N_iN_j}}\int_{-1}^{1} dx\int_0^{2\pi}d\phi_1d\phi_2~ \frac{d\chi_i(\bm{n})}{d\ln r}\chi_j(\bm{n}).
\end{equation}
Using the ansatz for $\chi_i(\bm{n})$ of \req{ansatz} and the Green's function in \req{Greenfunc}, we will find that the Berry connection matrix $\hat{D}$ is given by:
\begin{equation}
\label{Dij2}
D_{ij}=\frac{(\alpha_i\alpha_j+2\beta_i\beta_j)}{8\pi^2\sqrt{N_iN_j}(\nu_i-\nu_j)(\nu_i+\nu_j+1)},
\end{equation}
where the normalization factor $N_i$ of the angular eigenfunctions is calculated to be
\small
\begin{equation}
\label{normalization}
N_i=\frac{\Big{[} (\alpha^2_i+2\beta^2_i)\partial_{\nu_i}G_{\nu_i}(1)+[2\beta^2_i+4\alpha_i\beta_i]\partial_{\nu_i}G_{\nu_i}(-\frac{1}{2}) \Big{]}}{(4\pi)(2\nu_i+1)}.
\end{equation}
\normalsize
The details of the derivation of these results can be found in Appendix \ref{app:Dij}.  According to \req{deltau}, we have the correction to the effective potential of the lowest level as:
\begin{equation}
\label{corr}
\Delta u_0(r\rightarrow \infty)=\sum_{j\neq 0}|D_{0j}|^2.
\end{equation}
This can be calculated using the following trick. Firstly, \req{coeff} for the eigenstate coefficients $(\alpha,\beta)$ can be rewritten in a more compact form:
\begin{equation}
\hat{H}(\nu)\vec{\alpha}=0, ~~~ \vec{\alpha}=\begin{pmatrix}\alpha \\ \sqrt{2}\beta\end{pmatrix},
\end{equation}
where the $2\times2$ matrix Hamiltonian $\hat{H}(\nu)$ is
\small
\begin{equation}
\label{matrixH}
\hat{H}(\nu)=2\pi\begin{pmatrix}\frac{1}{\lambda_1}+G_{\nu}(1); & \sqrt{2}G_{\nu}\left(-\frac{1}{2}\right) \\ \sqrt{2}G_{\nu}\left(-\frac{1}{2}\right);  & G_{\nu}(1)+G_{\nu}\left(-\frac{1}{2}\right)-\frac{1}{\lambda_2}\end{pmatrix},
\end{equation}
\normalsize
and the normalization condition for the eigenstate coefficients $(\alpha,\beta)$ can be chosen as the following \footnote{The simple expression for the normalization condition is achieved by the fact that the matrix Hamiltonian is two-dimensional, real and symmetric, and that it has proper pole structures.}:
\begin{equation}
\label{normcon}
\vec{\alpha}\otimes\vec{\alpha}^T=\sigma_y\hat{H}\sigma_y=\det \hat{H}\cdot \hat{H}^{-1}
\end{equation}
Using the matrix Hamiltonian $\hat{H}(\nu)$ and the normalization condition defined above, we can express the righthand side of \req{corr} as a contour integration on the complex plane of variable $\nu$:
\begin{widetext}
\begin{equation}
\label{eq326}
\sum_{j\neq0}|D_{0j}|^2=\frac{1}{2}\left\{\frac{1}{2\pi i}\oint_{\mathcal{C}} d\nu\frac{Tr[\mathrm{Res}~\hat{K}(\nu_0)\cdot \hat{K}(\nu)]}{(\nu_0-\nu)^2}+\frac{1}{2\pi i}\oint_{\mathcal{C}} d\nu\frac{Tr[\mathrm{Res}~\hat{K}(\nu^*_0)\cdot \hat{K}(\nu)]}{(\nu_0^*-\nu)^2}\right\},
\end{equation}
\end{widetext}
where the matrix function $\hat{K}(\nu)$ is formally defined as $\hat{K}(\nu)=\hat{H}^{-1}(\nu)$. It has poles at where the matrix Hamiltonian has zeros, and decays rapidly enough when $|\nu|$ goes to infinity. The derivation of this result can be found in Appendix \ref{firstorder}, and the integration contour is shown in \reffig{contour}.

The integration contour can be deformed to enclose the other four poles off the real axis and the integration can be easily carried out, leading to the following result:
\begin{equation}
\Delta u_0=\sum_j|D_{0j}|^2=\frac{1}{3},
\end{equation}
which combined with \req{adiabatic} gives us the marginal result $\gamma=1$. This shows the importance of including Berry connection matrix $\hat{D}$ and the physical consistency. With this marginal situation, the existence and property of the three-particle bound state must be handled numerically.

\subsection{Zero Angular Momentum: Numerics}

\begin{figure}
\includegraphics[width=9cm]{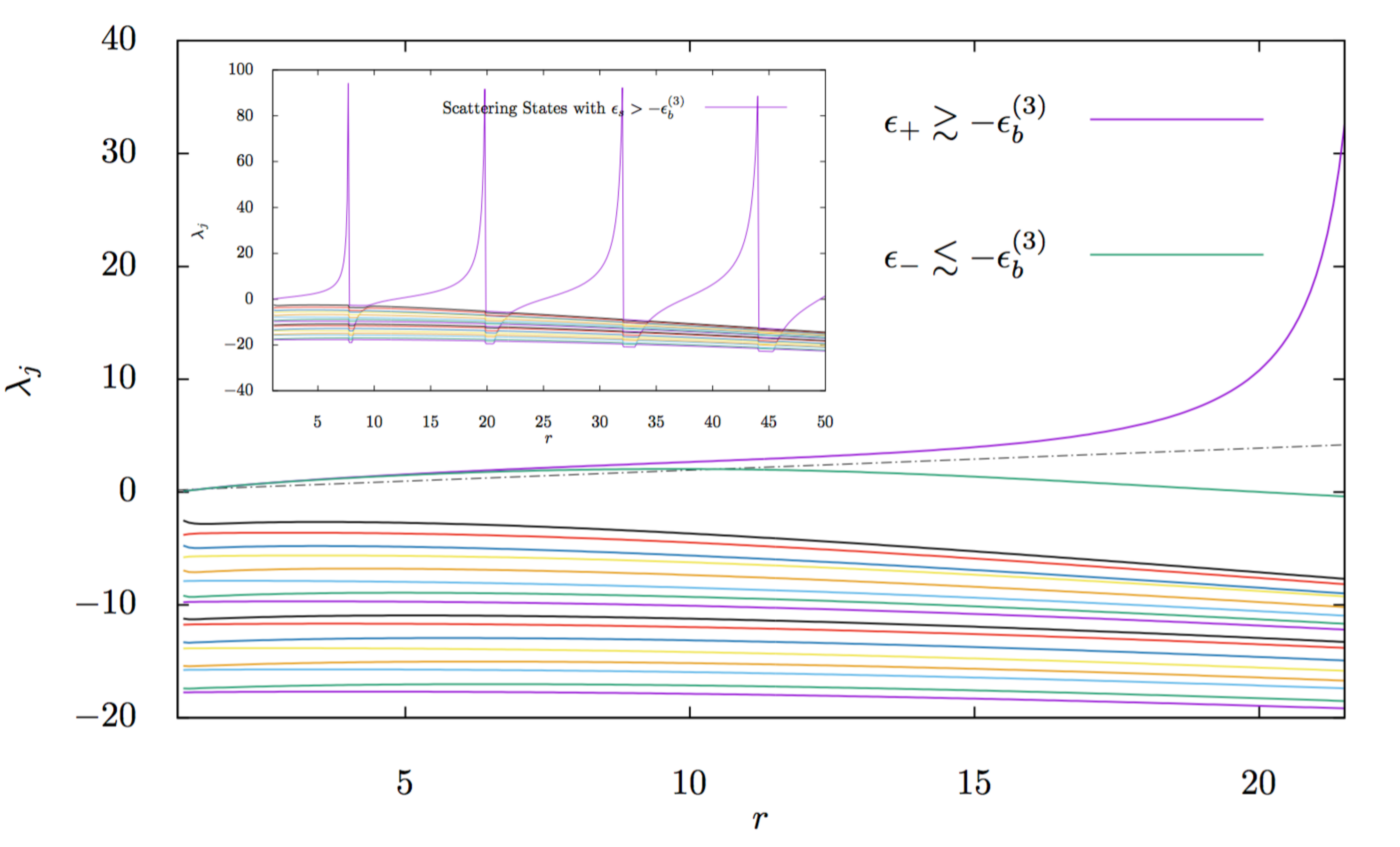}
\captionsetup{justification=raggedright}
\caption{\label{existence} Eigenvalues of matrix $\hat{\Lambda}$, calculated for energy slightly above ($\epsilon_+$) and below ($\epsilon_-$) the binding energy $\epsilon^{(3)}_b$. Inset shows result for energy well above $\epsilon^{(3)}_b$, which is the typical behavior for spherical waves.}
\end{figure}

The numerical implementation of the renormalization group equation (\ref{renorm2}) is simple, it is just a set of first-order ordinary differential equations and the second-order numerical integration algorithm is efficient enough for our purpose. The initial matrix \req{initialcon} is diagonal, the first-order correction matrix $\hat{D}$ is anti-symmetric and the effective potential matrix $\hat{u}$ is diagonal, these conditions guarantee that during the evolution all eigenvalues of matrix $\hat{\Lambda}$ are real as they should be. The algorithm is divided into two steps: firstly we run the renormalization process at energy slightly below the two-particle threshold, the existence of three-particle bound state is reflected in the divergence of the highest eigenvalue of $\hat{\Lambda}$ and the number of bound states equals to the number of jumps of the highest eigenvalue \footnote{Mathematically the divegence is positive on one side of the vertical asymptote and negative on the other side, thus there is jump from one side to the other side. These jumps are numerically realized by inverting the highest eigenvalue while keep the other eigenvalues intact when the former hits a sufficiently large value.} by Levinson's theorem \cite{Lin:1998pa,atomic}, as discussed at the end of Sec. \ref{sec:Levinson}. Secondly, if the bound state exists, we further run the renormalization process with varying energies to determine the binding energy of the three-particle bound state. Typical behaviors of different energies are shown in \reffig{existence}, where energy slightly above the three-particle binding energy shows a single jump and energy slightly below the three-particle binding energy shows no divergence. If the energy is well above the three-particle binding energy, the situation corresponds to a spherical wave, where periodic jumps will occur at the zeros of the wave function.

\begin{figure}
\includegraphics[height=5cm]{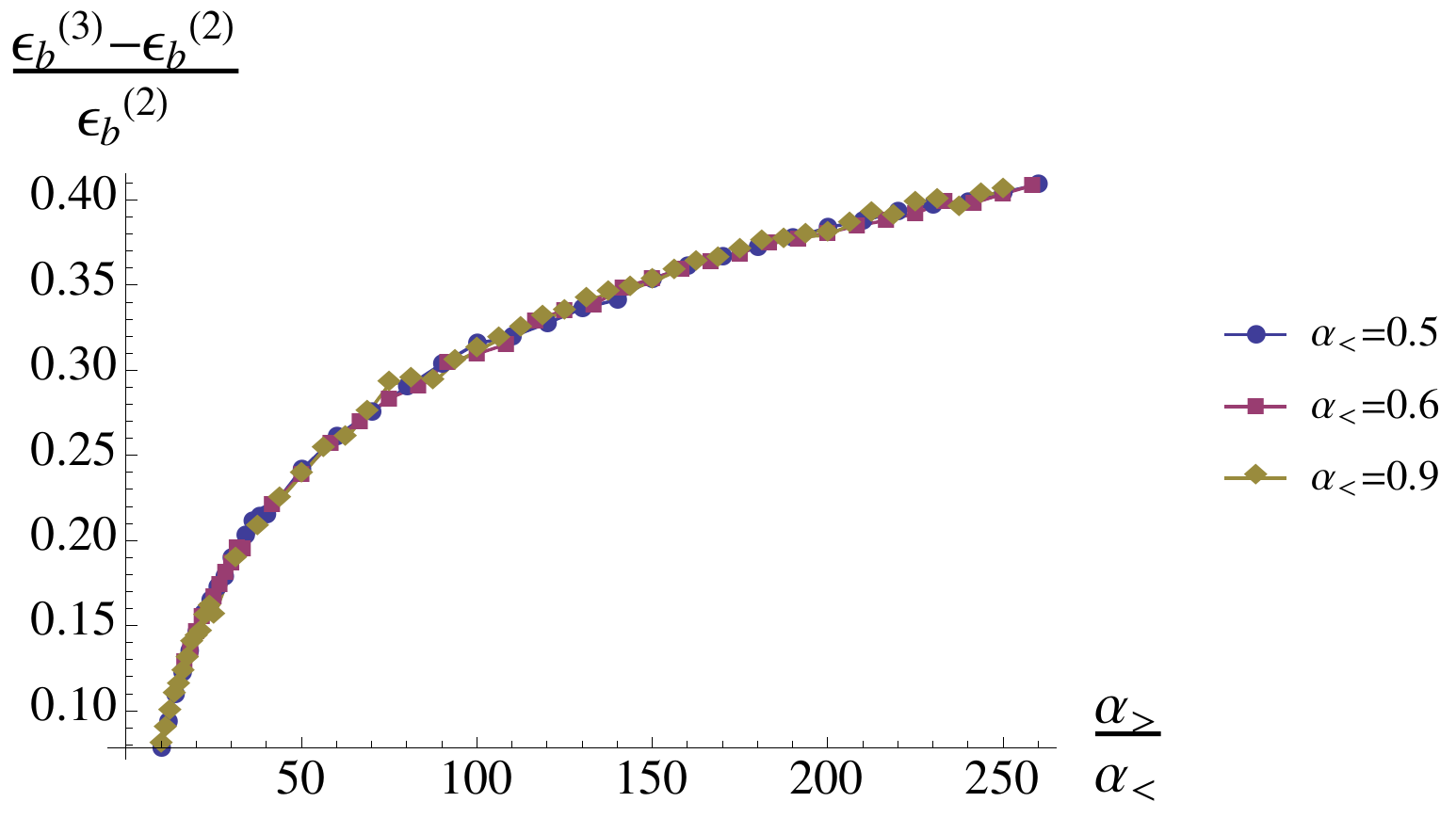}
\captionsetup{justification=raggedright}
\caption{\label{fig:limit} The univeral curve of $(\epsilon^{(3)}_b-\epsilon^{(2)}_b)/\epsilon^{(2)}_b$ versus $\alpha_>/\alpha_<$ at large scattering length ratios. Data points are collected in the region $\alpha_>/\alpha_<\geqslant 10$, and with three different values of $\alpha_<$. They fall on the same curve within the numerical accuracy.}
\end{figure}

The calculation is carried out using MATLAB \cite{MATLAB:2016} on a laptop with number of levels included $N=40$. Each run of the renormalization process takes less than 10 minutes \footnote{In the numerical calculation we need to carefully exclude the spurious level as discussed in the introduction section.} and inclusion of more levels only changes the result by less than 1\%. For zero angular momentum, there exists at most one three-particle bound state. At large $\alpha_>/\alpha_<$ ratio, the ratio between three-particle binding energy and the two-particle threshold versus $\alpha_>/\alpha_<$ falls on a universal curve, as illustrated in \reffig{fig:limit}. A similar universal curve also appears in the case of three-boson all interacting attractively \cite{0953-4075-44-20-205302,PhysRevA.85.025601,0953-4075-46-5-055301}. According to the result for vanishing intra-species interaction \cite{0953-4075-44-20-205302,PhysRevA.73.032724}, the universal curve in \reffig{fig:limit} should approach 1.39 asymptotically at infinite $\alpha_>/\alpha_<$ ratio. Curiously, the convergence to 1.39 is extremely slow: it only reaches 0.4 for $\alpha_>/\alpha_<=250$, the largest scattering length ratio shown in \reffig{fig:limit}. In fact, the curve reaches $\sim 1$ only for $\alpha_>/\alpha_<\sim 10^8$ and the correction to 1.39 in the large $\alpha_>/\alpha_<$ limit scales as $1/\ln (\alpha_>/\alpha_<)$. This curious fact can be partially understood from the first order perturbation theory with respect to the small parameter $f_<$ from \req{eq:amp}. It seems that the result $(\epsilon^{(3)}_b-\epsilon^{(2)}_b)/\epsilon^{(2)}_b=1.39$ is practically inaccessible due to the logarithmic slow convergence. Into the region with small $\alpha_>/\alpha_<$ ratio, universality breaks and the three-particle binding energy merges into the two-particle threshold at critical values, we listed several critical values in Table \ref{critical}. It's notable that our calculation only takes the two scattering lengths $\alpha_<$ and $\alpha_>$ as input parameters (see \req{spectrum1}). The microscopic cut-off $r_0$ only appears in the initial condition, where the kinetic energy dominates and the limit $r\rightarrow 0$ can be safely taken (see \req{initialcon}). These indicate that the property of the three-particle bound state depends only on the scattering lengths $\alpha_>,\alpha_<$, but not on the microscopic details of the interactions.

\begin{table}
\captionsetup{justification=raggedright}
\caption{\label{critical}Critical values of $\alpha_>$ corresponding to different $\alpha_<$ when the three-particle bound state disappear into the two-particle threshold.}
\begin{ruledtabular}
\begin{tabular}{l|rrrrrr}
$\alpha_<~~~$ & 0.80 & 0.82 & 0.85 & 0.90 & 0.95\\
\hline
$\alpha^c_>$ & 1.82 & 1.85 & 1.90 & 1.99 & 2.09
\end{tabular}
\end{ruledtabular}
\end{table}

\subsection{Non-Zero Angular Momentum}

The solution to \req{Green} with $\bm{n}'$ along north pole ($\bm{n}'=\bm{n}_1$) and total angular momentum $m\neq 0$ has the following form (see Appendix \ref{app:G}):
\small
\begin{equation}
\label{GreenGeneral}
\begin{split}
G^{(m)}_{\nu_j}(\bm{n},\bm{n}_1) =& (\bm{N}\cdot\bm{B_1})^{m}\frac{1}{4\cos \pi\left(\nu_j+\frac{1}{2}\right) }\frac{\Gamma(\nu_j+m+1)}{\Gamma(\nu_j+1)\Gamma(m+1)}\\
&\times ~R^{(m)}_{\nu_j}(1-\bm{N}^T\hat{A}_1\bm{N})\\
R^{(m)}_{\nu_j}(x)=& {_2F_1(-\nu_j,\nu_j+m+1;m+1;x)},
\end{split}
\end{equation}
\normalsize
where the four-dimensional vector $\bm{B}_1$ and $4\times4$ matrix $\hat{A}_1$ are defined as
\begin{equation}
A_1=\begin{pmatrix} 1 & & & \\ & 1 & & \\ & & 0 & \\& & & 0\end{pmatrix}, ~~~ \bm{B}_1=(0, 0, 1,  i)^T,
\end{equation}
and $\bm{N}$ is the following four-dimensional unit vector:
\begin{equation}
\bm{N}=
\begin{pmatrix}
\sqrt{\frac{1-x}{2}}\cos\phi' \\ \sqrt{\frac{1-x}{2}}\sin\phi' \\\sqrt{\frac{1+x}{2}}\cos\phi' \\ \sqrt{\frac{1+x}{2}}\sin\phi'
\end{pmatrix},
\end{equation}
where $\phi'$ is an arbitrary phase. To obtain the Green's function with $\bm{n}'$ along the other two poles, we rotate vector $\bm{B}_1$ and matrix $\hat{A}_1$ by $2\pi/3$ on three-dimensional unit sphere, which corresponds to $\pi/3$ rotation in four-dimensions. The rotation matrices are as follows:
\begin{equation}
\mathcal{R}_{2,3}=
\begin{pmatrix}
\frac{1}{2} & 0 & \mp\frac{\sqrt{3}}{2} & 0\\
0 & \frac{1}{2} & 0 & \mp\frac{\sqrt{3}}{2}\\
\pm\frac{\sqrt{3}}{2} & 0 & \frac{1}{2} & 0\\
0 & \pm \frac{\sqrt{3}}{2} & 0 & \frac{1}{2} \\
\end{pmatrix}.
\end{equation}
Applying the rotation matrices to  the four-dimensional vector $\bm{B}_1$ and $4\times4$ matrix $\hat{A}_1$ we get:
\small
\begin{equation}
\begin{split}
& \hat{A}_{2,3}=\mathcal{R}_{2,3}\hat{A}_1\mathcal{R}^{-1}_{2,3}=
\begin{pmatrix}
\frac{1}{4} & 0 & \pm\frac{\sqrt{3}}{4} & 0\\
0 & \frac{1}{4} & 0 & \pm\frac{\sqrt{3}}{4}\\
\pm\frac{\sqrt{3}}{4} & 0 & \frac{3}{4} & 0\\
0 & \pm\frac{\sqrt{3}}{4} & 0 & \frac{3}{4} 
\end{pmatrix},\\
& \bm{B}_{2,3}=\mathcal{R}_{2,3}\bm{B}_1=\left( \mp\frac{\sqrt{3}}{2} , \mp\frac{\sqrt{3}}{2}i , 1/2 ,i/2 \right)^T.
\end{split}
\end{equation}
\normalsize
Substituting the ansatz for eigenfunctions in \req{eigenstate1} with the above specification into \req{eigenstate2}, we will obtain the following constraints on the coefficients (here we add the superscript to emphasize the dependence on the angular momentum $m$):
\begin{equation}
\label{coeff2}
\begin{pmatrix} \frac{1}{\lambda_1}+G^{(m)}_{\nu_j}(11); & G^{(m)}_{\nu_j}(12)+G^{(m)}_{\nu_j}(13) \\ 
-G^{(m)}_{\nu_j}(21); & \frac{1}{\lambda_2}-G^{(m)}_{\nu_j}(22)-G^{(m)}_{\nu_j}(23) \end{pmatrix}\begin{pmatrix}\alpha_j \\ \beta_j\end{pmatrix}=0,
\end{equation}
where we have used the shortened notation $G^{(m)}_{\nu_j}(lm)\equiv G^{(m)}_{\nu_j}(\bm{n}_l,\bm{n}_m)$. Still the equation of spectrum is obtained via setting the determinant to zero. By performing the asymptotic analysis similar to those for zero angular momentum, we will obtain the following solution to the effective potential $u^{(m)}_0(r)$ up to first order correction (Appendix \ref{firstorder}):
\begin{equation}
u^{(m)}_0(r\rightarrow \infty)=-r^2\epsilon^{(2)}_b+(m^2-1)+O(r^{-2})
\end{equation}
thus for non-zero angular momentum, the wave function we will obtain is subject to power-law decay, and no three-particle bound state is guaranteed at large length scale. To confirm the absence of three-particle bound state, we need the calculation not only at large length scale, but also in the intermediate region, which we will still investigate numerically.

\begin{figure}
\includegraphics[scale=0.15]{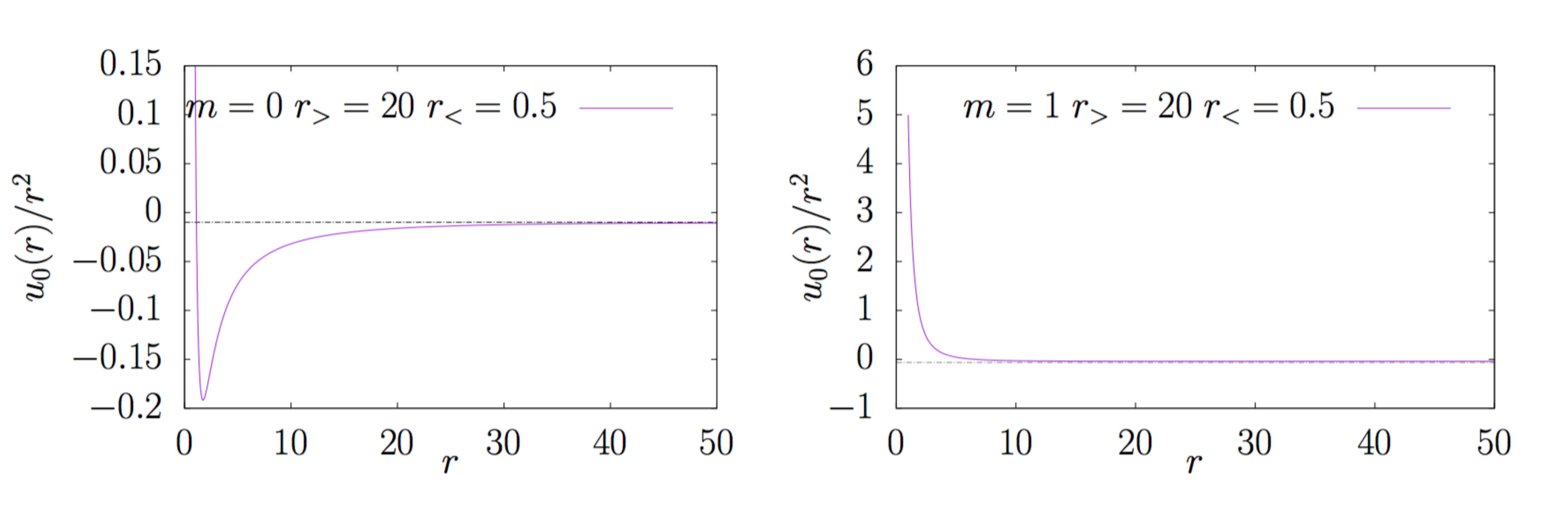}
\captionsetup{justification=raggedright}
\caption{\label{effpotential} Effective Potential $u_0(r)/r^2$ for $m=0,1$ with input parameters $\alpha_>=20$ and $\alpha_<=0.5$. The dashed line indicates the position of the two-particle threshold.}
\end{figure}

The numerical implementation for nonzero angular momentum is essentially the same as that for zero angular momentum, if we substitute the proper angular eigenfunctions into the corresponding formulas. Result shows that there is no three-particle bound state for nonzero angular momentum. To get a sense of what is happening among different $m$ values, we also calculated the effective potential $u_0(r)/r^2$ for the lowest level, the curve has minimum in case of $m=0$ while for $m>0$ the potential is monotonously decreasing with increasing $r$ (\reffig{effpotential}), then it is straightforward to see the possibility of getting three-particle system bounded for $m=0$ and its unlikeness for $m>0$.

\section{Conclusion}
\label{sec:conclusion}
In summary, we investigated the existence of three-particle bound states in a two-species, interacting bosonic system in two dimensions where coupling between like bosons is repulsive and otherwise attractive. We developed a simple and efficient algorithm via choice of proper parameterization and base functions. Large scale behavior of the system is handled analytically and interaction region is handled numerically. Our result shows that there is only one three-particle bound state for zero angular momentum, and it will merge into the two-particle threshold at small ratio between scattering lengths (the critical ratio $\alpha^c_>/\alpha_<$ is about $2.2\sim 2.3$). In contrast, there exist two three-particle bound states when the couplings between all the three bosons with equal masses are attractive, as investigated in the literature \cite{PhysRevA.19.425,0953-4075-44-20-205302,PhysRevA.85.025601,0953-4075-46-5-055301,PhysRevA.73.032724}. For non-zero angular momentum, there is no three-particle bound state. The two scattering lengths provide enough information to determine the three-particle binding energy, while the microscopic cut-off $r_0$ and the interaction constants $\lambda_{1,2}$ do not enter any way other than through the scattering lenghts. Our result is in agreement with the previous investigations \cite{0953-4075-44-20-205302,PhysRevX.4.031020} in the sense that there are only finite number of three-particle bound states in two dimensions, in contrast to the condensation of infinite number of three-particle bound states in three dimensions, and we showed this fact both analytically (the parameter $\gamma$ define in \req{gamma} is equal to or smaller than unity, which excludes the possibility of infinite number of bound states) and numerically.

Existing approaches for this kind of quantum three-body problem in the literature are mainly different variations of the Skorniakov-Ter-Martirosian methods. It can be implemented in real space and solved via the integral equations for the scattering amplitude \cite{PhysRevA.67.010703,PhysRevLett.93.090404}; or be implemented in momentum space and solved via the diagrammatic techniques for scattering matrix \cite{PhysRevA.73.032724,PhysRevA.73.053607,PhysRevA.81.043634}. It can also be converted into a series of solvable differential equations \cite{PhysRevA.72.060702}. All these approaches involve several numerical integrations over unbounded spaces or kernel inversion, some of them are limited to s-wave resonant scattering. The hyperspherical method has been used for few-body problems in two dimensions \cite{Nielsen1999,Nielsen2001373,PhysRevA.74.042506,PhysRevLett.112.103201,PhysRevA.90.042707,PhysRevA.91.062710}, with the usual approach of representing states as a sum of many hyperspherical harmonics. Our paper therefore provides an alternative approach to the quantum three-body problems, simple and efficient, involving only direct root finding and evolving of a first-order ordinary differential equation to an intermediate length scale (for example, the divergence behavior showing the existence of bound state is already clear at a relatively small length scale $r\sim 20$ in \reffig{existence}, and there is no need to evolve the equation further to any larger length scale). It is capable of handling both short- and long- range physics, free of numerical instability and converges fast enough to avoid parallelism on clusters. Also our choice of basis via Hopf coordinates reduces the squared proliferation of hyperspherical harmonics to a linear one with increasing number of included levels, which saves greatly in numerical endeavor. 

We believe that the scenario considered here is realized experimentally in the excitonic systems in two dimensions (for example, the GaAs-based quantum well structures), and it is natural to generalize the present formalism to four-particle problems or to fermionic systems. Further investigation of four-particle problem should reveal more quantitative features of the two-component bosonic systems in two dimensions, thus shedding more light on the rich phenomena observed in excitonic system in quantum well structures or microcavities. Another system that the present results can be potentially applied to is the bosonic dipoles in the bi-layer geometry, where dipoles on the same layer attract \cite{PhysRevA.80.052702} while on different layers repel \cite{PhysRevA.82.044701,PhysRevA.83.043602}, and the short-range limit is applicable for sufficiently large interlayer distances or small dipole lengths. The present formalism only considered bosonic systems in two dimensions, where three-particle states only exist in s-wave channel and there are only a finite number of them. It is recently proposed that the fermionic system in two dimensions fine-tuned to p-wave resonance can host infinite tower of three-particle bound states, which is called the super Efimov effect \cite{PhysRevLett.110.235301}. Since our formalism can handle all possible scattering channels by construction, it is promising to generalize the present formalism to the fermionic systems to verify the existence of the proposed super Efimov effect in p-wave channel.

\begin{acknowledgments}
We would like to thank Yuri Rubo, Rui Hu and Zimo Sun for helpful discussions, and Leonid Glazman for valuable comments on the manuscript. This work is supported by Simons foundation.
\end{acknowledgments}

\appendix
\numberwithin{equation}{section}

\section{Hopf Coordinates, Contact Interactions and Laplacian}
\label{app:Hopf}
Under the Hopf Coordinates in \req{Hopf} and the three-dimensional unit vectors defined in \req{unit}, we can express the distances between particles as follows:
\begin{equation}
|\bm{r}_{12}|^2=\frac{r^2}{2}(1-\bm{n}\cdot\bm{n}_1), ~~~ |\bm{r}_{i3}|^2=\frac{r^2}{2}(1-\bm{n}\cdot\bm{n}_{i+1}), ~~~ i=1,2.
\end{equation}
We make the choice that the interaction between particle 3 with the other two is attractive and the interaction between particle 1 and 2 is repulsive.
\begin{equation}
V_{12}=\lambda_1\delta^2(\bm{r}_{12}), ~~~ V_{i3}=-\lambda_2\delta^2(\bm{r}_{i3}), ~~~ i=1,2.
\end{equation}
The short-ranged interactions are modeled as contact interaction with finite radius $r_0$, thus the $\delta$ function in the above expression actually depends on length scale in the following manner:
\small
\begin{equation}
\begin{split}
\delta^2(\bm{r})\rightarrow& \frac{1}{\pi}\delta\Big{[}\frac{r^2}{2}(1-\bm{n}\cdot\bm{n}')-r^2_0\Big{]}\\
& =\frac{2}{\pi r^2}\delta\Big{[} (1-\bm{n}\cdot\bm{n}')-\frac{2r^2_0}{r^2} \Big{]}\equiv\frac{2}{\pi r^2}\delta_r(1-\bm{n}\cdot\bm{n}')
\end{split}
\end{equation}
\normalsize
This particular form of the cut-off is not unique, but only observable values of $\alpha_{>,<}$ enter into the final result.

Under the Hopf coordinates, the full Laplacian operator can be calculated using the covariant form
\begin{equation}
\nabla^2=\frac{1}{\sqrt{g}}\nabla_i\sqrt{g}g^{ij}\nabla_j, ~~~ g=\det\hat{g},
\end{equation}
where Hopf variables are $(r,x,\phi_1,\phi_2)$ and the metric is
\begin{equation}
\hat{g}=\begin{pmatrix} 1 & & & \\ & \frac{r^2}{4(1-x^2)} & & \\ & & \frac{(1-x)r^2}{2} & \\ & & & \frac{(1+x)r^2}{2} \end{pmatrix}.
\end{equation}
The result is just what we got in the main text:
\begin{equation}
-\nabla^2=-\frac{1}{r^3}\frac{\partial}{\partial r}r^3\frac{\partial}{\partial r}+\frac{4\hat{L}^2}{r^2},
\end{equation}
where the angular momentum operator is
\begin{equation}
\hat{L}^2=-\frac{\partial}{\partial x}(1-x^2)\frac{\partial}{\partial x}-\frac{\partial^2_{\phi_1}}{2(1-x)}-\frac{\partial^2_{\phi_2}}{2(1+x)}.
\end{equation}
Correspondingly, the separation of variable for an angular function $F(x,\phi_1,\phi_2)$ with desired symmetry is
\begin{equation}
F(x,\phi_1,\phi_2)=f(x)e^{im_1\phi_1+im_2\phi_2},
\end{equation}
then the eigenstates are labeled by the quantum number set $(l,m_1,m_2)$, where $l(l+1)$ is the eigenvalue of operator $\hat{L}^2$, and $m_{1,2}$ are integer numbers. By including interaction terms, we replace the operator $4\hat{L}^2$ with the effective potential operator $\hat{U}(r)=4\hat{L}^2+r^2\hat{V}(r)$ in \req{eq:effpotential}. Consequently we replace the quantum number $l$ with effective potential $u(r)$, where $u(r)$ is the eigenvalue of operator $\hat{U}(r)$, while keeping the quantum numbers $m_{1,2}$ intact. This separation of variable scheme in accordance with Hopf coordinates enables us to consider different angular momentum $m_{1,2}$ separately, and just as what we got in the main text, only the sector with zero angular momentum $m_1=m_2=0$ hosts the possible bound state.

\begin{figure}
\includegraphics[scale=0.15]{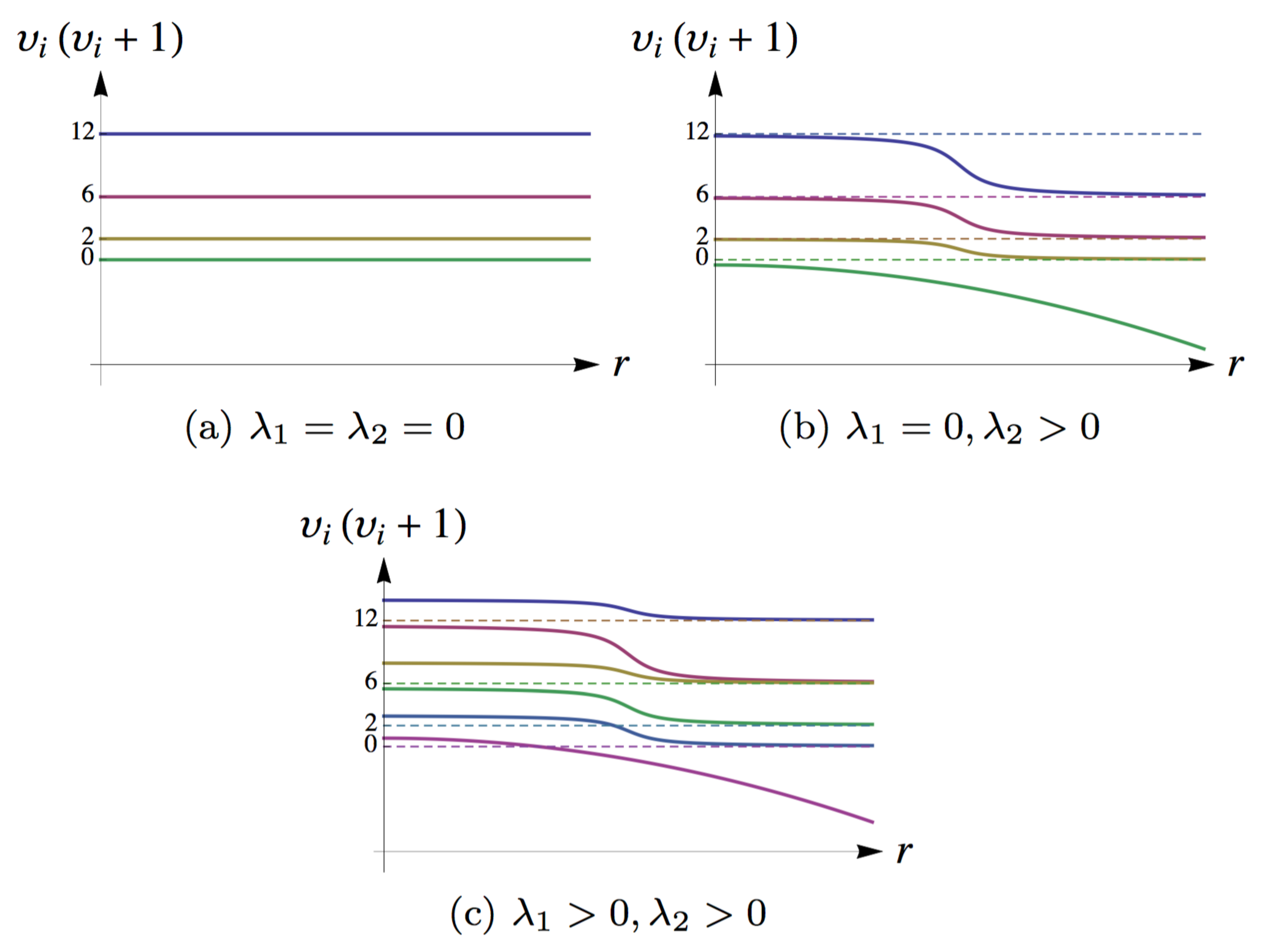}
\captionsetup{justification=raggedright}
\caption{\label{fig:alter} Schematic diagram for eigenstates of several low-lying levels within the zero angular momentum sector, where $u_i=4\nu_i(\nu_i+1)$ is the eigenvalue of the effective potential operator $\hat{U}(r)$. In (a) with $\nu_i=l_i$, it is just the eigenvalue of angular momentum operator $\hat{L}^2$; In (b), only attraction is included and only one state is altered for each level, the other unaltered states are denoted as dashed lines; In (c), both attraction and repulsion is included, and unaltered states are still denoted as dashed lines.} 
\end{figure}

Within the zero angular momentum sector, we have argued that there are at most two relevant states for each level in the main text, which is shown in \reffig{fig:alter}. This leads to the conclusion that in search of possible bound state, we only need to consider at most $2N$ states, where $N$ is the number of levels included.

\section{Green's Function}
\label{app:G}
Following the definition in \req{Green}, we first solve for the north pole $\bm{n}'=\bm{n}_1$ and then rotate the solution to the other two poles. The construction of the Green's function can be carried out following the standard procedure of separation of variables:
\begin{equation}
\begin{split}
& G(\bm{n},\bm{n}_1)=G(x)\exp(im_1\phi_1+im_2\phi_2),\\
& \hat{L}^2 =\left[-\frac{\partial}{\partial x}(1-x^2)\frac{\partial}{\partial x}+\frac{m_1^2}{2(1-x)}+\frac{m_2^2}{2(1+x)}\right].
\end{split}
\end{equation}
In expansion of Green's function in terms of eigenfunctions of $\hat{L}^2$, we only need to consider those that are connected to the $\delta$-function, thus we require $m_1=0$ and the eigenfunctions to be regular around $x=-1$. These eigenfunctions then can be represented in terms of hypergeometric functions \cite{Tables}:
\small
\begin{equation}
\label{angulareigen}
\begin{split}
& \hat{L}^2 X^{(m)}_j(x)=\left(j+\frac{m}{2}\right)\left(j+\frac{m}{2}+1\right)X^{(m)}_j(x),\\
& X^{(m)}_j(x)=\left(\frac{1+x}{2} \right)^{m/2}R^{(m)}_{j}\left(\frac{1+x}{2}\right),\\
& R^{(m)}_{j}(x)={_2F_1}(-j,j+m+1;m+1;x),
\end{split}
\end{equation}
\normalsize
where $m=m_1+m_2=m_2$ and $j$ takes the value $0,1,2,\cdots$. By using the representation of $\delta$-function
\small
\begin{equation}
\delta(x)=\frac{1}{2}\sum_{j=0}^{\infty}\frac{(-)^j(2j+m+1)\Gamma(j+m+1)}{\Gamma(m+1)\Gamma(j+1)}X^{(m)}_j(x),
\end{equation}
\normalsize
we immediately obtain the expression for Green's function:
\small
\begin{equation}
\begin{split}
G^{(m)}_{\nu}(\bm{n},\bm{n}_1) =& (\bm{N}\cdot\bm{B_1})^{m}\frac{1}{4\cos \pi(\nu+\frac{1}{2}) }\frac{\Gamma(\nu+m+1)}{\Gamma(\nu+1)\Gamma(m+1)}\\
&\times ~R^{(m)}_{\nu}(1-\bm{N}^T\hat{A}_1\bm{N}),
\end{split}
\end{equation}
\normalsize
where the vector $\bm{B}_1$ and matrix $\hat{A}_1$ are
\begin{equation}
A_1=\begin{pmatrix} 1 & & & \\ & 1 & & \\ & & 0 & \\& & & 0\end{pmatrix}, ~~~ \bm{B}_1=(0, 0, 1,  i)^T.
\end{equation}

\section{Equation of Spectrum and Asymptotic Analysis - General Angular Momentum}
\label{app:C}
The equation of spectrum is obtained by setting the determinant of \req{coeff2} to zero. In order to calculate the involved quantities $G^{(m)}_{\nu_j}(lm)$, we need to put the three-dimensional unit vectors in \req{unit} back on the four-dimensional unit sphere. This can be done using the following correspondence:
\begin{equation}
\begin{split}
& \bm{n}_1\rightarrow \bm{N}_1=(0,0,\cos\phi',\sin\phi')^T,\\
& \bm{n}_{2,3}\rightarrow \bm{N}_{2,3}=(\mp\frac{\sqrt{3}}{2}\cos\phi',\mp\frac{\sqrt{3}}{2}\sin\phi',\frac{1}{2}\cos\phi',\frac{1}{2}\sin\phi')^T,
\end{split}
\end{equation}
where $\phi'$ is an arbitrary phase. By direct calculation we will obtain the following results
\begin{equation}
\begin{split}
& \begin{cases}
G^{(m)}_{\nu}(11)=e^{im\phi'}f(\nu,m)R^{(m)}_{\nu}(1)\\
G^{(m)}_{\nu}(12)=\frac{1}{2^m}e^{im\phi'}f(\nu,m)R^{(m)}_{\nu}(\frac{1}{4})\\
G^{(m)}_{\nu}(13)=\frac{1}{2^m}e^{im\phi'}f(\nu,m)R^{(m)}_{\nu}(\frac{1}{4})
\end{cases},\\
& \begin{cases}
G^{(m)}_{\nu}(21)=\frac{1}{2^m}e^{im\phi'}f(\nu,m)R^{(m)}_{\nu}(\frac{1}{4})\\
G^{(m)}_{\nu}(22)=e^{im\phi'}f(\nu,m)R^{(m)}_{\nu}(1)\\
G^{(m)}_{\nu}(23)=\frac{(-)^m}{2^m}e^{im\phi'}f(\nu,m)R^{(m)}_{\nu}(\frac{1}{4})
\end{cases},
\end{split}
\end{equation}
where the factor $f(\nu,m)$ is defined as
\begin{equation}
f(\nu,m)=\frac{1}{4\cos \left[\left(\nu+\frac{1}{2}\right)\pi\right] }\frac{\Gamma(\nu+m+1)}{\Gamma(\nu+1)\Gamma(m+1)},
\end{equation}
and $R^{(m)}_{\nu}(x)$ has a singularity at $x=1$ which is regularized by the finite radius $r_0$:
\small
\begin{equation*}
f(\nu,m)R^{(m)}_{\nu}(1)=\frac{1}{4\pi }\left[\ln\frac{16}{\delta}-\Psi(-\nu)-\Psi(\nu+m+1)+2\Psi\left(\frac{1}{2}\right)\right],
\end{equation*}
\normalsize
where $\delta=r^2_0/r^2$ and the $\Psi(x)$ is the digamma function \cite{Tables}. Putting all these results together, we finally obtain the equation of spectrum for general value of $m$:
\begin{widetext}
\begin{equation}
\label{spectrumm}
\Big{[}\ln\frac{r}{\alpha_<}-\frac{1}{2}M(\nu_j,m) \Big{]}\Big{[}\ln\frac{r}{\alpha_>}-\frac{1}{2}M(\nu_j,m)+2\pi (-)^mN(\nu_j,m) \Big{]}=2\Big{[}2\pi N(\nu_j,m)\Big{]}^2,
\end{equation}
\end{widetext}
with the following definition of the relevant quantities:
\small
\begin{equation}
\begin{split}
& M(\nu,m)=\Psi(-\nu)+\Psi(\nu+m+1),\\
& N(\nu,m)=\frac{1}{2^m}\frac{1}{4\cos \left[\left(\nu+\frac{1}{2}\right)\pi\right] }\frac{\Gamma(\nu+m+1)}{\Gamma(\nu+1)\Gamma(m+1)}R^{(m)}_{\nu}\left(\frac{1}{4}\right),
\end{split}
\end{equation}
\normalsize
where ${\mathbb{C}}=0.577\cdots$ is the Euler constant. Specification of \req{spectrumm} to the case $m=0$ is just what we got previously in \req{spectrum1}.

We then analyze the large scale behavior of the lowest level. With increasing length scale $r$, the angular eigenvalue $u_0$ becomes more and more negative, and the imaginary part of $\nu_0$ becomes larger. In the limit $|u_0|=-u_0\rightarrow \infty$, we have the following asymptotic behavior of the relevant functions \cite{Tables}:
\begin{equation}
\begin{split}
& M(\nu_0,m) \sim \ln |u_0|-2\ln 2-\frac{4-3m^2}{3|u_0|},\\
& N(\nu_0,m) \sim \frac{\exp(-\frac{2\pi}{3}\sqrt{|u_0|})}{|u_0|^{1/4}}.
\end{split}
\end{equation}
Then asymptotically, the equation of spectrum \ref{spectrumm} reduces to
\begin{equation}
\ln\frac{r}{\alpha_>}= \frac{1}{2}\ln |u_0|-\ln 2-\frac{4-3m^2}{6|u_0|},
\end{equation}
where only the solution associated with $\alpha_>$ is chosen because the other solution associated with $\alpha_<$ corresponds to the spurious state discussed previously in the introduction section.

Solving this equation iteratively we will get the large scale behavior of the effective potential:
\begin{equation}
u^{(m)}_0(r\rightarrow \infty)=-r^2\epsilon^{(2)}_b+(3m^2-4)/3+O(r^{-2}),
\end{equation}
where $\epsilon^{(2)}_b=4/\alpha^2_>$ is the two-particle threshold binding energy. This is the result under adiabatic approximation.

\section{Calculation of the Matrix Elements of the Berry Connection}
\label{app:Dij}
Here we calculate the matrix elements of the Berry connection for the case with zero angular momentum $m=0$. Firstly we verify the orthogonality of the eigenstates $\chi_i(\bm{n})$ in \req{ansatz} and calculate the normalization factor $N_i$ in \req{normalization} via the overlap integral:
\begin{equation}
\label{eqnD1}
\braket{\chi_i}{\chi_j}=\frac{1}{8\pi^2}\int_{-1}^{1} dx\int_0^{2\pi}d\phi_1d\phi_2~ \chi_i(\bm{n})\chi_j(\bm{n}).
\end{equation}
Substituting the expression in \req{ansatz} into the above integral, we will get
\begin{equation}
\braket{\chi_i}{\chi_j}=(\alpha_i\alpha_j+ 2\beta_i\beta_j) I_1+2(\alpha_i\beta_j+\alpha_j\beta_i+\beta_i\beta_j)I_2,
\end{equation}
where the two integral $I_{1,2}$ are
\begin{equation}
\label{eqnD3}
\begin{split}
& I_1\equiv \frac{1}{8\pi^2}\int_{-1}^{1}dx\int_0^{2\pi}d\phi_1d\phi_2 ~ G_{\nu_i}(\bm{n}\cdot\bm{n}_1)G_{\nu_j}(\bm{n}\cdot\bm{n}_1),\\
& I_2\equiv \frac{1}{8\pi^2}\int_{-1}^{1}dx\int_0^{2\pi}d\phi_1\phi_2 ~ G_{\nu_i}(\bm{n}\cdot\bm{n}_1)G_{\nu_j}(\bm{n}\cdot\bm{n}_2).
\end{split}
\end{equation}
Substituting \req{Greenfunc} into \req{eqnD3} and performing the integration, we will get
\begin{equation}
\begin{split}
& I_1=\frac{1}{4\pi}\frac{1}{(\nu_i-\nu_j)(\nu_i+\nu_j+1)}\left[G_{\nu_i}(x)-G_{\nu_j}(x)\right]_{x\rightarrow 1},\\
& I_2=\frac{1}{4\pi}\frac{1}{(\nu_i-\nu_j)(\nu_i+\nu_j+1)}\left[G_{\nu_i}(x)-G_{\nu_j}(x)\right]_{x=-\frac{1}{2}}.
\end{split}
\end{equation}
Substitute these back into the overlap integral in \req{eqnD1}, and apply the constraint on coefficients $(\alpha_i,\beta_i)$ in \req{coeff}, we finally arrived at
\small
\begin{equation}
\braket{\chi_i}{\chi_j}=\begin{cases}
0 ~~~~~~~~~~~~~~~~~~~~~~~~~~~~~~~~~~~~~~~~~~~~~~~~~~~~~~ (i\neq j)\\
\frac{\Big{[} (\alpha^2_i+2\beta^2_i)\partial_{\nu_i}G_{\nu_i}(1)+[2\beta^2_i+4\alpha_i\beta_i]\partial_{\nu_i}G_{\nu_i}\left(-\frac{1}{2}\right) \Big{]}}{(4\pi)(2\nu_i+1)} ~~~ (i=j)
\end{cases}
\end{equation}
\normalsize

Secondly we try to calculate the matrix element of the Berry connection in \req{Dij}. The scale dependence only appears in the regularized Green's function $G_{\nu_i}(1)$, thus the relevant quantity that contributes to the derivative with respect to $\ln r$ is inside integral $I_1$:
\begin{equation}
I_1=\frac{1}{4\pi} \frac{1}{(\nu_i-\nu_j)((\nu_i+\nu_j+1)}\frac{1}{2\pi}\ln \frac{r_i}{r_j}+(\cdots),
\end{equation}
where we have used the expression for $G_{\nu_i}(1)$ in \req{reg}, and the scale $r$ is equipped with subscript to differentiate between $\ket{\chi_i}$ and $\ket{\chi_j}$. The derivative involved in \req{Dij} is then performed with respect to $r_i$ and we finally get the expression for the matrix element $D_{ij}$ as
\begin{equation}
D_{ij}=\frac{(\alpha_i\alpha_j+2\beta_i\beta_j)}{8\pi^2\sqrt{N_iN_j}(\nu_i-\nu_j)(\nu_i+\nu_j+1)},
\end{equation}
with the normalization factor $N_i$ given by $\braket{\chi_i}{\chi_i}$.

\section{First Order Correction to Adiabatics}
\label{firstorder}
In this section we present the detail of calculation of the first order correction to the effective potential $u_0$ in \req{corr} and obtain the result in \req{eq326}. Using the matrix Hamiltonian $\hat{H}(\nu)$ in \req{matrixH} and normalization condition in \req{normcon}, together with the expression for the first order correction in \req{Dij2}, we arrive at the following expression:
\small
\begin{equation}
\begin{split}
|D_{0j}|^2&=\frac{(2\nu_0+1)(2\nu_j+1)}{[\nu_0(\nu_0+1)-\nu_j(\nu_j+1)]^2}\frac{Tr[\vec{\alpha}_0\cdot\vec{\alpha}_0^T\cdot\vec{\alpha}_j\cdot\vec{\alpha}_j^T]}{\partial_{\nu_0}\det\hat{H}~\partial_{\nu_j}\det\hat{H}}\\
& =\frac{(2\nu_0+1)(2\nu_j+1)}{[\nu_0(\nu_0+1)-\nu_j(\nu_j+1)]^2}\frac{Tr[\det \hat{H}_0\cdot\hat{H}^{-1}_0\det \hat{H}_j\cdot\hat{H}^{-1}_j]}{\partial_{\nu_0}\det\hat{H}~\partial_{\nu_j}\det\hat{H}}
\end{split}
\end{equation}
\normalsize
The last line can be further simplified to the following form using the residuals of $\hat{K}=\hat{H}^{-1}$:
\begin{equation}
|D_{0j}|^2=\frac{(2\nu_0+1)(2\nu_j+1)}{[\nu_0(\nu_0+1)-\nu_j(\nu_j+1)]^2}Tr[\mathrm{Res}~\hat{K}(\nu_0)\cdot \mathrm{Res}~\hat{K}(\nu_j)].
\end{equation}
The total correction $\Delta u_0$ in \req{corr} can then be converted into a contour integration on the complex plane of variable $\nu$, where the integration contour is along the real axis. On that complex plane, each $\nu_j$ is a first-order pole along the positive real axis, and each have its counterpart on the negative real axis. There are four extra poles off real axis, corresponding to true bound state ($\nu_0$) and spurious bound state ($\nu_s$) respectively (see \reffig{contour}). Finally the expression for $\Delta u_0$ are as follows:
\begin{widetext}
\begin{equation}
\begin{split}
\sum_{j\neq0}|D_{0j}|^2=&\sum_{j\neq 0}\frac{(2\nu_0+1)(2\nu_j+1)}{[\nu_0(\nu_0+1)-\nu_j(\nu_j+1)]^2}Tr[\mathrm{Res}~\hat{K}(\nu_0)\cdot \mathrm{Res}~\hat{K}(\nu_j)]\\
=&\sum_{j\neq 0}\frac{\partial}{\partial \nu_0}\Big{[}-\frac{1}{\nu_0-\nu_j}+\frac{1}{\nu_0+\nu_j+1}\Big{]}Tr[\mathrm{Res}~\hat{K}(\nu_0)\cdot \mathrm{Res}~\hat{K}(\nu_j)]\\
=&\frac{1}{2}\left\{\frac{1}{2\pi i}\oint_{\mathcal{C}} d\nu\frac{Tr[\mathrm{Res}~\hat{K}(\nu_0)\cdot \hat{K}(\nu)]}{(\nu_0-\nu)^2}-\frac{1}{2\pi i}\oint_{\mathcal{C}} d\nu\frac{Tr[\mathrm{Res}~\hat{K}(\nu_0)\cdot \hat{K}(\nu)]}{(\nu_0+\nu+1)^2}\right\}\\
=&\frac{1}{2}\left\{\frac{1}{2\pi i}\oint_{\mathcal{C}} d\nu\frac{Tr[\mathrm{Res}~\hat{K}(\nu_0)\cdot \hat{K}(\nu)]}{(\nu_0-\nu)^2}+\frac{1}{2\pi i}\oint_{\mathcal{C}} d\nu\frac{Tr[\mathrm{Res}~\hat{K}(\nu^*_0)\cdot \hat{K}(\nu)]}{(\nu_0^*-\nu)^2}\right\},
\end{split}
\end{equation}
\end{widetext}
where we have used the fact that $\mathrm{Res}~\hat{K}(\nu_0)=-\mathrm{Res}~\hat{K}(\nu_0^*)$ and the factor $1/2$ appears because we are only summing over positive real poles. This is nothing but \req{eq326} in the main text.

Similarly we can calculate the same first order correction to adiabatics for the case of non-zero angular momentum $m$. By direct calculation similar to the case of zero angular momentum we obtain the following result for the Berry connection $D_{ij}$:
\small
\begin{equation}
\begin{split}
D_{ij}&=\frac{1}{8\pi^2\sqrt{N_iN_j}}\int_{-1}^{1} dx\int_0^{2\pi}d\phi_1d\phi_2~ \frac{d\chi^{(m)}_i(\bm{n})}{d\ln r}\chi^{(m)}_j(\bm{n})\\
& =\frac{1}{8\pi^2\sqrt{N_iN_j}}(\alpha_i\alpha_j+2\beta_i\beta_j)\frac{1}{2^m}\frac{1}{(\nu_i-\nu_j)(\nu_i+\nu_j+m+1)},
\end{split}
\end{equation}
\normalsize
where the normalization factor $N_i$ of the angular eigenfunctions is calculated to be
\small
\begin{equation}
N_i=\frac{\Big{[} (\alpha^2_i+2\beta^2_i)\partial_{\nu_i}G^{(m)}_{\nu_i}(1)+[2(-)^m\beta^2_i+4\alpha_i\beta_i]\partial_{\nu_i}G^{(m)}_{\nu_i}(-\frac{1}{2}) \Big{]}}{2^m(4\pi)(2\nu_i+m+1)},
\end{equation}
\normalsize
and the single-argument Green's function is defined as
\begin{equation}
G^{(m)}_{\nu}(x)=\left(\frac{1+x}{2}\right)^{m/2}f(\nu,m)R^{(m)}_{\nu}\left(\frac{1+x}{2}\right),
\end{equation}
where the functions $f(\nu,m)$ and $R^{(m)}_{\nu}(x)$ are defined in Appendix \ref{app:C}. For the correction to the effective potential of the lowest level, we still have
\begin{equation}
\Delta u^{(m)}_0(r\rightarrow \infty)=\sum_{j\neq 0}|D_{0j}|^2.
\end{equation}
This can be calculated exactly the same way as the case $m=0$. The only change is the expression for the matrix Hamiltonian $\hat{H}(\nu)$:
\small
\begin{equation}
\hat{H}(\nu)=2\pi\begin{pmatrix}\frac{1}{\lambda_1}+G^{(m)}_{\nu}(1); & \sqrt{2}G^{(m)}_{\nu}\left(-\frac{1}{2}\right) \\ \sqrt{2}G^{(m)}_{\nu}\left(-\frac{1}{2}\right);  & G^{(m)}_{\nu}(1)+(-)^mG^{(m)}_{\nu}\left(-\frac{1}{2}\right)-\frac{1}{\lambda_2}\end{pmatrix}.
\end{equation}
\normalsize
The rest of the calculation is essentially the same as the case $m=0$, and the contour integration trick  eventually gives us the following result:
\begin{equation}
\begin{split}
& \Delta u^{(m)}_0(r\rightarrow \infty)=1/3,\\
& u^{(m)}_0(r\rightarrow \infty)=-r^2\epsilon^{(2)}_b+(m^2-1)+O(r^{-2}).
\end{split}
\end{equation}
This shows that the marginal value $\gamma=1$ is only realized when $m=0$; for non-zero angular momentum, no bound state is guaranteed for the system.

\bibliography{ThreeBody}

\end{document}